\documentstyle[pre,aps,epsf]{revtex}
      \newcommand{\beq}{\begin{equation}}
      \newcommand{\eeq}{\end{equation}}
      \newcommand{\beqa}{\begin{eqnarray}}
      \newcommand{\eeqa}{\end{eqnarray}}
      
      \newcommand{\nn}{\nonumber}
      \newcommand{\Tr}{{\rm Tr}}

      \newcommand{\del}{\partial}

      \newcommand{\al}{\alpha}
      \newcommand{\be}{\beta}
      \newcommand{\ga}{\gamma}
      \newcommand{\de}{\delta}

      \newcommand{\ba}{\bbox{a}}
\newcommand{\bc}{\bbox{c}}
\newcommand{\bee}{\bbox{e}}
\newcommand{\bg}{\bbox{g}}

\newcommand{\bV}{\bbox{V}}
\newcommand{\bL}{\bbox{L}}
\newcommand{\bbe}{\bbox{\beta}}

\begin{document}
\draft

%\begin{flushright}
%NUP-A-97-20\\
%SMC-PHYS-151\\
%cond-mat/9711308\\
%November, 1997
%\end{flushright}
%
\title{Renormalization group for renormalization-group equations toward
 the universality classification 
of infinite-order phase transitions} 

\author{Chigak Itoi\thanks{\tt e-mail: itoi@phys.cst.nihon-u.ac.jp} 
\thanks{On leave from Department of Physics, 
College of Science and Technology, 
Nihon University, Kanda Surugadai, Chiyoda-ku, 
Tokyo 101-8308, Japan}}
\address{Department of physics and astronomy, 
University of British Columbia, Vancouver, BC, V6T1Z1, Canada }

\author{Hisamitsu Mukaida
\thanks{\tt e-mail: mukaida@saitama-med.ac.jp} }
\address{Department of Physics, Saitama Medical College,  
Kawakado, Moroyama, Saitama, 350-0496, Japan} 

\maketitle
                
\begin{abstract}
We derive a new renormalization group to calculate the nontrivial critical 
exponent of the divergent correlation length thereby giving a universality 
classification of essential singularities in infinite-order phase 
transitions. This method thus resolves the vanishing scaling 
matrix problem. The exponent is obtained from the maximal eigenvalue of a 
scaling matrix in this renormalization group, as in the case of ordinary 
second-order phase transitions.  We exhibit several nontrivial 
universality classes in infinite-order transitions different from the 
well-known Berezinski\u\i-Kosterlitz-Thouless transition.
\end{abstract}

\pacs{05.70.Fh, 05.70.Jk, 11.10.Hi, 64.60.Ak} 
%{\small {\it Keywords:} BKT transition, essential singularity, 
%renormalization
%group, marginal perturbation, universality class }}}
%
\section{Introduction}
The Berezinski\u\i-Kosterlitz-Thouless (BKT) transition 
is well-known as an infinite-order phase transition \cite{BKT}.
The correlation length $\xi$ has an essential singularity at the 
critical coupling parameter $g_c$
\begin{equation}
\xi \sim \exp ( A |g-g_c|^{-\sigma}  ),
\label{xi}
\end{equation} 
with the critical exponent $\sigma = 1/2$ or 1. In $c=1$ conformal
 field theory(CFT)\cite{BZ},
there are infinitely many models where infinite-order phase 
transitions can occur.
Any of them shows the same universality as the BKT transition.

One observes 
$\sigma$ different from 1/2 or 1 in some $c >1$ CFTs \cite{IK,YNH,B}.   
Recently, a model of a quantum spin chain,  whose long-distance behavior 
is described by level 1 $SU(N)$ WZW model, 
% with a critical exponent $\sigma= \frac{N}{N+2}$, 
%where $N > 1$ is an integer, 
was studied by Itoi and 
Kato \cite{IK}. They pointed out 
that an infinite-order phase transition with 
a critical exponent $\sigma= \frac{N}{N+2}$
occurs by an $SU(N)$ symmetry-breaking
marginal operator. In the $N=3$ case, this corresponds to the 
gapless Haldane gap
phase transition in a spin 1 isotropic antiferromagnet 
in one dimension. 
In a problem of dislocation-mediated melting,
some curious numbers were observed by Young, Nelson and Halperin \cite{YNH}.
They obtained $\sigma=1/2$ for a model on a square lattice, 
$\sigma=2/5$
for a simplified model on 
a triangular lattice and a non-algebraic number $\sigma = 0.36963...$
for a generalized model. 
In Ref.\cite{B}, Bulgadaev studied topological 
phase transitions in $c > 1$ CFT with non-abelian symmetry, where 
non-abelian vortices play an important role. 
 They belong to special classes of infinite-order 
phase transitions
and several series of  
$\sigma$ dependent on the symmetry of the system, were found.  
Though there have been several studies for those some different types of
models with infinite-order phase transitions including BKT type, 
the universality classification 
by this critical exponent still remains a challenging problem. 

In this paper, we study the universal nature of 
the critical exponent $\sigma$ in infinite-order 
phase transitions.
We show that the critical exponent
$\sigma$ is determined from the operator product coefficients 
of the marginal operators which cause the infinite-order phase
 transition.
It is shown that a marginally irrelevant operator
can also affect the value of the critical exponent $\sigma$.

%In general, we need to know long-distance asymptotic behavior 
%of running coupling constants for obtaining a critical exponent 
%of the correlation length. It is determined by
%the beta function near a fixed point. 
The critical exponent of the correlation length is extracted from 
a long-distance asymptotic  form of running coupling 
constants, whose leading term is determined by 
the motion of the coupling 
constants near a fixed point. 
In an ordinary finite-order 
phase transition, we linearize the renormalization-group equation 
(RGE) around the fixed point 
and can derive the exponent exactly. Namely, we can show that 
the inverse of the exponent is equal to the maximal eigenvalue of 
the scaling matrix defined by the derivative of the beta function 
at the fixed point.  
%In an ordinary finite-order phase transition, 
%we obtain a critical point as a zero of the beta function
%in a renormalization group equation (RGE) 
%and obtain the critical exponent of the correlation length by
%the inverse of the maximal eigenvalue of the scaling 
%matrix given by the derivative of the beta function.
One does not have to solve the differential equation exactly
in order to obtain the exact critical exponents in this case.
In the infinite-order phase transition however, 
the scaling matrix vanishes
since the phase transition is driven only by 
marginal operators. 
So far, one has had to solve the differential equations
explicitly to obtain the critical exponent, 
although   RGEs 
with multiple variables are generally non-integrable 
due to their non linearity
except for some fortunate cases like the BKT transition.
This difficulty is one of the reasons why
the universality classification of infinite-order phase transitions
by the critical exponent $\sigma$ in eq.(\ref{xi})
has never been successfully done. 

In order to resolve this problem, 
we apply another renormalization group (RG) method developed 
in Refs.\cite{BK,CGO}
to studying the long-distance asymptotic behavior of the solution of the 
original non-integrable RGE.
This RG method is starting to be recognized as a general tool 
for asymptotic analysis.   
Chen, Goldenfeld and Oono\cite{CGO} introduced the idea of RG to 
singular perturbation theory and gave a unified treatment.  
According to Bricmont, Kupiainen and Lin\cite{BK}, 
the RG transformation for a partial differential equation is defined as 
a semi-group transformation on a space of initial data, which is generated 
by a scaling transformation combined with time evolution. 
%Here, we employ a new renormalization group (RG) method 
%to study the long-distance asymptotic behavior of the solution 
%of the original non-integrable RGE. 
%Recently, there are remarkable studies of the long 
%time asymptotic behavior of nonlinear differential 
%equations by an RG method\cite{BK,CGO}.  According to Ref.\cite{BK}, 
%RG transformation for a partial differential equation is defined as 
%a semi-group transformation on a space of initial data, which 
%is generated by a scaling transformation combined with time evolution. 
%%which is applicable to
%%many problems of non-linear equations \cite{BK,CGO}.
Koike, Hara and Adachi used this general method practically in the
study of the
critical phenomenon in the Einstein equation of the gravitational
collapse with formation of black holes \cite{KHA}. Tasaki 
gave a pedagogical example for the RG transformation, where 
the equations of motion in Newtonian gravity were analyzed\cite{T}.

In Sec.\ref{rgeforrge},  
we reuse the RG transformation of Ref.\cite{T}, which
enables us to calculate the critical exponent $\sigma$ in 
eq.(\ref{xi}) without solving the nonlinear differential 
equation explicitly. 
The new RGE regards the straight flow line solving the original RGE
as a fixed point, where
the derivative of the beta function in the new RGE  
has non-zero value in general.
 In Sec.\ref{massive}, we show that the inverse of the maximal 
eigenvalue of the scaling matrix derived from the new RGE 
gives the critical exponent $\sigma$.  In Sec.\ref{massless},  
we also study asymptotic behavior of the running coupling constants in a 
massless phase and 
extend the well-known formula for a logarithmic finite-size correction 
to the case of multiple running 
coupling constants.  In Sec.\ref{examples}, we exhibit 
several nontrivial examples motivated by 
antiferromagnetic quantum spin chains.  
Finally, we give a summary and discussions in Sec.\ref{discussion}. 

\section{RGE for RGE}
\label{rgeforrge}
\subsection{Formalism}
\label{formalism}
Let us begin with the RGE for a given set of $n$ marginal operators
\begin{equation}
\frac{d \bg}{d t} = \bV (\bg),  
\label{orge2}
\end{equation}
where  $\bg =(g_1, \cdots, g_n)$ is a set of coupling parameters and 
$t= \log l$ with $l$ being a length scale parameter. 
Since the operators are all marginal,  the right-hand side is 
expanded as 
\begin{equation}
V_k(\bg) = \sum_{ij} C_k ^{ij} g_i g_j+{\rm O}(g^3),
\label{gigj}
\end{equation}
where $C_k^{ij}$ is proportional to the operator product coefficients 
of the operators. First we neglect the higher-order terms ${\rm O}(g^3)$, 
and later we discuss  
irrelevance of those neglected terms.   

In general, we find several critical surfaces where the RG flow
is absorbed into the origin.  A phase transition 
occurs if the initial coupling constants cross one of the critical
surfaces. 
These critical surfaces divide the coupling parameter space into 
several regions which are phases. 
In the next section, we consider one massive phase 
surrounded by a set of critical
surfaces, where there are several marginally relevant
 coupling parameters.
In this region, we have a finite correlation length, 
which becomes larger as the coupling parameter 
approaches the critical surface.

\begin{figure}[ht]
\setlength{\unitlength}{1mm}
\begin{picture}(150, 55)(-40,0)
        \put(15,5){\epsfysize=5cm \epsfbox{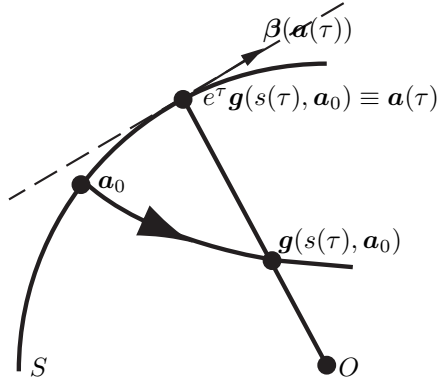}}
        \put(59,5){$O$}
        \put(18,5){$S$}
        \put(51,22){$\bg(s(\tau),\ba_0)$}
        \put(27,30){$\ba_0$}
        \put(49, 50){$\bbox{\beta}(\ba(\tau))$}
        \put(41,41){$e^{\tau} \bg(s(\tau),\ba_0) \equiv \ba(\tau)$}
\end{picture}
\caption{
Illustration for ${\cal R}_\tau$ and the beta function defined in eq.(\ref{rgrg}). 
For simplicity, we take $n=2$. The dashed line represents the 
tangent space at $\ba(\tau) \in S$. 
}
\label{R_tau}
\end{figure}

We are going to study the long-distance asymptotic behavior of solutions 
for the RGE (\ref{orge2}) which is non-integrable in general. 
To this end, let us introduce the RG method explained in the introduction. 
We define a renormalization group transformation on  $n-1$ 
dimensional sphere which forms a set of initial values. 
We denote the solution $\bg$ 
of eq.(\ref{orge2}) with the initial condition
$\ba_0 =(a_{01}, \cdots, a_{0 n})$  as
\begin{equation}
\bg(t, \ba_0),
\label{inita}
\end{equation}
namely, $\bg(0, \ba_0) = \ba_0$. 
The function $ e^{\tau}\bg(e^{\tau} t, \ba_0)$
is a solution of the RGE (\ref{orge2}) as well, 
because of its scale invariance. 
Let $S$ be the $n-1$ dimensional sphere whose center is at the origin 
with the radius $|\ba_0| \equiv a_0$. 
We define a new renormalization-group
transformation ${\cal R}_{\tau}: S \rightarrow S$ 
as follows: 
\begin{equation}
{\cal R}_{\tau} \ba_0
\equiv e^{\tau} \bg (s(\tau), \ba_0) \equiv \ba(\tau).   
\label{rgt}
\end{equation}
Note that ${\cal R}_\tau$ has a semi-group property: 
\beq
        {\cal R}_{\tau_1 + \tau_2} = {\cal R}_{\tau_2} \circ {\cal R}_{\tau_1}.
\eeq
The meaning of ${\cal R}_\tau$ is the following: first,  choose $\tau$. 
Then  move $\ba_0$ along the solution $\bg(t, \ba_0)$ during the time 
$s(\tau)$. 
Here $s(\tau)$ is determined by the condition  
$\bg(s(\tau), \ba_0) e^{\tau} \in S$. See 
Fig.\ref{R_tau}. 
%%%%%%%%%%%Fig: R_tau-test%%%%

%%%%%%%%%%%%%end of Fig: R_tau-test%%%%%%%%%%%%

Next let us derive the beta function for ${\cal R}_{\tau}$.  
Noting that $\bV(\bg)$ is quadratic,   we have 
\beqa
        \frac{d \ba}{d \tau} &=& 
        \ba + e^{\tau} \bV(\bg(s, \ba_0)) \frac{ds}{d\tau}\nn\\
        &=& \ba + e^{-\tau} \bV(\ba) \frac{ds}{d\tau}.   
\label{da/dtau}
\eeqa
The length-preserving condition 
\begin{equation}
        \ba \cdot \frac{d \ba}{d \tau}  = 0
\label{orth}
\end{equation}
leads to the following differential equation for $s(\tau)$:
\begin{equation}
\frac{ds}{d\tau} =   - \frac{e^{\tau} a_0^2}
{\ba \cdot \bV(\ba)}
\label{s} 
\end{equation}
with the initial condition $s(0)=0$.
Inserting eq.(\ref{s}) into eq.(\ref{da/dtau}), we obtain the beta function
for ${\cal R}_\tau$.   
\begin{equation}
\beta_i (\ba) \equiv \frac{d a_i}{d \tau} =  
\frac{a_i \ba\cdot\bV(\ba)-V_i(\ba) a_0^2}
{\ba \cdot \bV (\ba)}.  
\label{rgrg}
\end{equation}

Note that $\bbe$ can be written as 
\beq
        \bbe(\ba) = 
        -\frac{a_0^2}{\ba \cdot \bV (\ba)} P(\ba)
\bV(\ba), 
\label{cfree}
\eeq
where $P$ is the $n \times n$ matrix that 
projects $\bV(\ba(\tau))$ onto the tangent space at $\ba(\tau) \in S$: 
\beq
        P_{i j}(\ba)  \equiv 
        \de_{i j} - \frac{a_i a_j}{a_0^2}.
\eeq 

For later use, we derive a polar coordinates representation of
the new RGE.  Employing polar coordinates, 
$\ba \in S$ is expressed as 
\beq
        \ba = \left(a_0 \prod_{\al=1}^{n-1} \sin \theta_\al, \, 
        a_0 \cos \theta_1 
        \prod_{\al =2}^{n-1}\sin \theta_\al, \, 
        a_0 \cos \theta_2 \prod_{\al =3}^{n-1}\sin \theta_\al, \, 
        \cdots, \, a_0 \cos \theta_{n-1} \right). 
        \eeq
Since 
$\{ \del \ba/\del \theta_\al \}_{1\leq \al \leq n-1}$ are orthogonal 
to each other, we 
can make the basis 
$\{ \tilde{\bee}_\al \}_{\al} $ orthonormal
on the tangent space at $\ba \in S$ by an appropriate rescaling 
\beq
        \tilde{\bee}_\al \equiv f_{\al} (\ba)^{-1} \frac{\del \ba}
        {\del \theta_\al}, 
        \ \ f_\al (\ba) \equiv 
        \left|  \frac{\del \ba}{\del \theta_\al} \right|.  
\label{tildee}
\eeq
Then, 
\beq
        \tilde\be_\al \equiv  \bbe \cdot \tilde{\bee}_\al  = 
        \frac{d \ba}{d \tau} \cdot \tilde{\bee}_\al  = 
        f_\al (\ba) \frac{d \theta_\al}{d \tau}, 
\label{tildebe}
\eeq
which leads to the RGE in polar coordinates
\beq
         \frac{d \theta_\al}{d \tau}= 
        \left( f_\al (\ba) \right)^{-1} \tilde\be_\al(\ba). 
        \label{polar}
\eeq

Returning to the coordinate-free representation eq.(\ref{cfree}), 
let us find a fixed point of the new RGE (\ref{rgrg}).
The nature of the new RGE near the fixed point
determines the  universal behavior of the 
infinite-order phase transition, as in the ordinary RGE 
for a finite-order one. 
Near  a fixed point, $\ba(\tau)$ moves slower 
as  its trajectory 
tends to a critical surface. 
This implies that the time $\ba(\tau)$ spent in a neighborhood of 
the fixed point is a singular function of the initial condition
$\ba_0$.   This singularity can occur only at fixed points of the new RGE,
which allows us to analyze its singular behavior by a linearization 
near the fixed points.
      
    From eq.(\ref{cfree}), one finds that 
$\ba$ is definitely a 
fixed point
if  $P(\ba) \bV(\ba) = {\bf 0}$ and
 $\ba \cdot \bV(\ba) \neq 0$.  
  In this case, since $\bV(k \ba)$ is parallel to $\ba$ for all 
real numbers $k$,  $\ba$ is on a straight flow line of the original 
RGE (\ref{orge2}).    Straight flow lines are put into two classes. 
If an arbitrary point $\ba$ on a straight flow line satisfies 
$\ba \cdot \bV(\ba) < 0$,  it is said to be an {\sl incoming straight 
flow line} because $\ba$ is carried toward the origin in time evolution. 
On the other hand,  if  $\ba \cdot \bV(\ba) > 0$ for all $\ba $ on a straight 
flow line, it is called an {\sl outgoing straight flow line}. 
If a fixed point $\ba$ 
of eq.(\ref{rgrg}) is on an incoming straight flow line, 
$-\ba$ is a fixed point on an outgoing straight flow line. 

What happens if $P(\ba) \bV(\ba)=\ba \cdot \bV(\ba) =0$ ?  
In this case, 
$\bV(\ba)$ itself vanishes.  It means that $\ba$ is a fixed point 
of the original RGE (\ref{orge2}).  Moreover, since $\bV$ is homogeneous, 
$k \ba$  is also a fixed point for all $k \in {\bf R}$.  
Namely, the original RGE (\ref{orge2}) has a fixed line in this case. 
If the original RGE (\ref{orge2}) has this fixed line,  
a point on the fixed line 
has a non-vanishing scaling matrix even though 
the coupling constants are all marginal at the trivial fixed 
point $\bg = {\bf 0}$.  Therefore, we can directly analyze the 
original RGE near a point on the fixed line and 
can show that the phase transition generally becomes 
of finite order in this case. 

\begin{figure}[ht]
\setlength{\unitlength}{1mm}
 \begin{picture}(150, 55)(-40,0)
        \put(15,5){\epsfysize=5cm \epsfbox{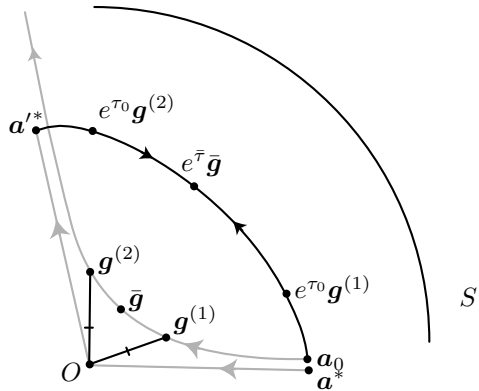}}
        \put(54,7){$\ba_0$}
        \put(73,15){$S$}
        \put(54,4){$\ba^*$}
        \put(51,16){$e^{\tau_0}\bg^{(1)}$}
        \put(35,12){$\bg^{(1)}$}
        \put(36,33){$e^{\bar{\tau}}\bar{\bg}$}
        \put(29,15){$\bar{\bg}$}
        \put(25,20){$\bg^{(2)}$}
        \put(25,40){$e^{\tau_0}\bg^{(2)}$}
        \put(20,5){$O$}
        \put(13,38){${\ba'}^*$}
\end{picture}
\caption{
        An example in the case of a flow having a turning point.  
Here we take $n=3$.  
The gray lines represent 
solutions of the original RGE (\ref{orge2}), 
while black ones on $S$ for the new RGE (\ref{rgrg}). 
Here $\ba^*$(${\ba'}^*$)  on the 
incoming(outgoing) straight 
flow line is a fixed point of the new RGE. 
}
\label{global}
\end{figure}

Here, we give a couple of remarks on 
the global nature of the new RG transformation 
${\cal R}_\tau$ defined by
(\ref{rgt}).   
First, there could be a turning point $\bar{\bg}$   
where 
$\bV(\bar{\bg})\cdot \bar{\bg} = 0$  with
$\bV(\bar{\bg}) \neq \bbox{0}$. Let 
\beq
        \log  \frac{a_0}{\left| \bar{\bg} \right|} \equiv \bar \tau. 
\label{bartau}  
\eeq
Although eqs.(\ref{s}) and (\ref{rgrg}) cannot be  defined  
at $\tau = \bar{\tau}$,
it is obvious in a geometric sense that  $\ba(\bar{\tau})$ and 
$s(\bar{\tau})$ are well-defined. For example, in Fig.\ref{global},  
$\ba(\bar\tau) = e^{\bar{\tau}} \bar{\bg}$ and $s(\bar{\tau})$ are 
 determined as the definite time 
$\bg(t, \ba_0)$ spent during the journey from $\ba_0$ to $\bar{\bg}$.  
%%%%%%%%%%%%%%%%%%%%
%Fig: global
%%%%%%%%%%%%%%%%%%%%
%%%%%%%%%%%%%%%%%%%%%%
%End of Fig: global
%%%%%%%%%%%%%%%%%%%%%%

Second,  if $\bg(t, \ba_0)$ has turning points,
$\ba(\tau)$ and $s(\tau)$ become multi-valued with respect to $\tau$. 
For example, in Fig.\ref{global}, 
$\bg(t, \ba_0)$ has a turning point at 
$\bar{\bg}$.  Suppose that $| \bg^{(1)} | = |\bg^{(2)}| = b$ and 
choose $\tau = \log a/b \equiv \tau_0$.  Then 
${\cal R}_{\tau_0}\ba_0$ has two images, $e^{\tau_0} \bg^{(1)}$ and  
$e^{\tau_0} \bg^{(2)}$.  
In this case we distinguish the images 
as $\ba^{(1)}(\tau_0)$ and $\ba^{(2)}(\tau_0)$.  
Similarly, $s(\tau_0)$ also has the same multiplicity, which is 
distinguished in a similar way. 
In Fig.\ref{global}, 
the image of ${\cal R}_\tau$ starting at $\ba_0$ reaches  
a branch point $e^{\bar{\tau}}\bar{\bg}$.  
We denote the solution from $\ba_0$ to 
$e^{\bar{\tau}}\bar{\bg}$  by $\ba^{(1)}(\tau)$.  
The remaining part is called $\ba^{(2)}(\tau)$.  
Both $\ba^{(1)}(\tau)$ and $\ba^{(2)}(\tau)$ are absorbed into 
the branch point $\bar{\bg}e^{\bar{\tau}}$, which indicates the fact that 
$\bar{\bg}$ gives a minimum distance from  the origin.  If a turning 
point corresponds to a maximum distance, the two solutions 
will escape from the branch point. 

%Note that a flow by ${\cal R}_\tau$ 
%changes its direction at a turning point, 
%as depicted in Fig.\cite{global}. If a turning point gives 
%maximum distance from the origin,  the corresponding point on 
%$S$ absorbs the flows in eq.(\ref{rgrg}).  In contrast, the flows escape 
%from it if the turning point has minimum distance. 

\subsection{Example -- the  2D classical XY model}
\begin{figure}[ht]
\setlength{\unitlength}{1mm}
 \begin{picture}(150, 50)(-10,0)
        \put(0,5){\epsfxsize=15cm \epsfbox{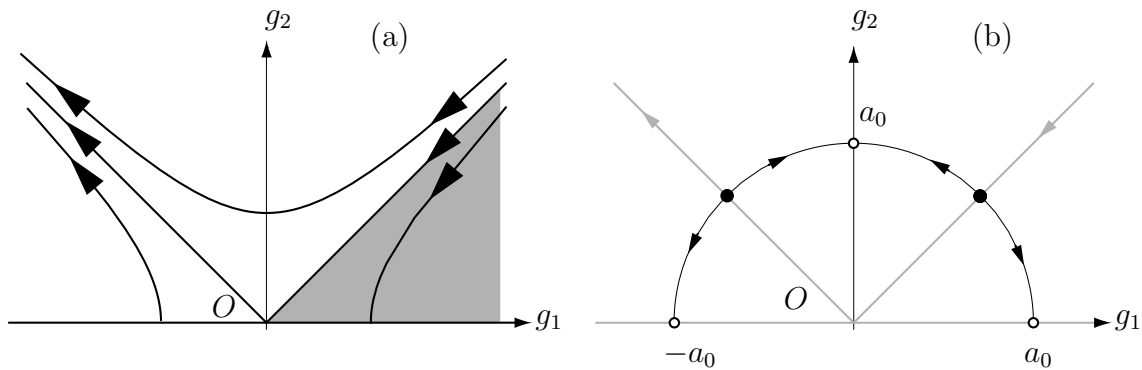}}
%       \put(65,5){\epsfxsize=5cm \epsfbox{xyflows2.eps}}
        \put(50,45){\large (a)}
        \put(130,45){\large (b) }
        \put(105,10){\large $O$}
        \put(29,9){\large $O$}
        \put(36, 48){\large $g_2$}
        \put(114, 48){\large $g_2$}
        \put(72,8){\large $g_1$}
        \put(149, 8){\large $g_1$}
        \put(89, 3){\large $-a_0$}
        \put(137, 3){\large $a_0$}
        \put(115, 35){\large $a_0$}
\end{picture}
\caption{
(a) Flow generated by the original RGE (\ref{xy}) of the XY model.  
(b) Flow of the new RGE (\ref{rgrg}) with 
eq.(\ref{xy}).  Fixed points are denoted by the black circles.  The new 
RGE is not defined at the white circles which correspond to 
a fixed line or turning points in Fig.\ref{xyflows}(a).  
}
\label{xyflows}
\end{figure}

Here we exhibit the new RGE for the 2D classical XY model as an 
illustrative example.  The original RGE for the XY model is given 
by\cite{KW}
\beqa
        \left(
        \begin{array}{c}
                \frac{d g_1}{d t}  \\
                \frac{d g_2}{d t} 
        \end{array}
        \right) 
        = \bV(\bg) \equiv
        \left(
        \begin{array}{c}
        -  g_2^2 \\
        -g_1 g_2 
        \end{array}
        \right)
\label{xy}
\eeqa
with $g_2 \geq 0$.  
Let us first look at the phase structure from eq.(\ref{xy}).  
See Fig.\ref{xyflows}(a). 
%
%%%%%%%%%%%%%% 
%   Fig: xyflows
%%%%%%%%%%%%%%
%%%%%%%%%%%%%end of Fig: xyflows%%%%%%%%%%%%

The RGE (\ref{xy}) has two straight 
flow lines $g_1 \pm g_2=0$   
and one fixed line $g_2 =0$.  It is well known that 
each point on the fixed line corresponds to the theory of a 
2D massless free boson that is parametrized by a compactification radius 
of the boson field\cite{BZ}.  The shaded region in Fig.\ref{xyflows}(a) 
is a massless phase since flow in this region is finally absorbed 
into a point on the $g_2 =0$ fixed line.  The incoming straight flow line,  
$g_1-g_2= 0$ with $g_2 \geq 0$ forms the phase boundary. 
  As an initial coupling approaches the phase boundary from the 
massive phase,  the correlation length $\xi$ becomes 
divergent.\footnote{
The correlation length also diverges when the initial 
coupling constants tend to a fixed point on 
$g_2 =0$ with $g_1 < 0$.  However,  the scaling matrix at this point 
 does not vanish and the ordinary finite-order phase transition
 takes place.  Since our interest is focused on an infinite-order
 phase transition, we do not consider that case here. 
}  

Now we turn to the new RGE for the XY model,  which is given by 
eqs.(\ref{rgrg}) and (\ref{xy}) with the condition 
$a_1^2 + a_2^2 = a_0^2$ ($a_2 \geq 0$).  
It is explicitly represented as 
\beq
        {\frac{d a_1}{d \tau} \choose \frac{d a_2}{d \tau}} = 
        {\frac{1}{2 a_1}\left(a_1^2 -a_2^2\right)
         \choose \frac{1}{2 a_2}\left( a_2^2 - a_1^2 \right)}
\label{nRGExy}  
\eeq
in cartesian coordinates.  Alternatively, using polar coordinates 
$\ba = (a_0 \sin \theta, \, a_0 \cos \theta)$ 
$(-\pi/2 \leq \theta \leq \pi/2)$, 
the RGE  becomes 
\beq
        \frac{d \theta}{d \tau} = - \cot 2\theta
\label{polarxy}
\eeq
owing to eq.(\ref{polar}).  

Next we find fixed points of the new RGE.   Solving 
\beq
        P(\ba) \bV(\ba) = \frac{1}{a_0^2} 
        \left( 
        \begin{array}{c}
        a_2^2 (a_1^2 - a^2_2) \\
        - a_1 a_2 (a_1^2 - a^2_2)
        \end{array}
        \right)
        = \bbox{0}, 
\eeq 
we have $\ba = (\pm a_0/\sqrt{2},  a_0/\sqrt{2}),  (\pm a_0, 0)$. 
Evaluating $\bV(\ba) \cdot \ba$ at those points, it turns out that 
$(a_0/\sqrt{2}, a_0/\sqrt{2})$ is on an incoming straight flow line 
while $(a_0/\sqrt{2}, - a_0/\sqrt{2})$  on an 
outgoing straight flow line. 
The remaining points $(\pm a_0, 0)$ are found to be on a fixed line. 
Note that $\bV(\ba) \neq \bbox{0}$ and $\bV(\ba) \cdot \ba = 0$ when 
$\ba = (0, a_0)$.  This means that 
$(0, a_0)$ corresponds to turning points on trajectories 
generated by the original RGE. 
The flow of the new RGE cannot be defined at the points $(0, a_0)$ and 
$(\pm a_0, 0)$.  Since the example is a simple two-parameter system, 
we can understand qualitative aspects of the global flow 
in the new RGE.  
See  Fig.\ref{xyflows}(b). 

In the last part of the next section, we will continue the analysis of
this model and derive $\sigma$ from the beta function
in eq.(\ref{nRGExy}).  
Before performing that, we need to know a representation of the correlation 
length in terms of the new RGE.

\section{Critical exponent of the correlation length in a massive phase}
\label{massive}
In this section, we explain how to evaluate the critical
exponent $\sigma$ of the correlation length in a massive phase
from the beta function (\ref{rgrg}).

We first define a correlation length $\xi(\ba_0)$ 
by the following formula: 
\beq
        |\, \bg(\log \xi(\ba_0), \ba_0) \, | =1. 
\label{xi2}
\eeq
Namely, $\log \xi(\ba_0)$ is the time 
$\bg(t, \ba_0)$ spent in the perturbative region.
We note that  $\xi(\ba_0)$ defined above changes as 
\beq
        {\rm e}^t \,  \xi (\bg(t, \ba_0)) = \xi (\ba_0)  
\eeq
under the original RG transformation,  
which should be satisfied by  an intrinsic length scale of the model.
The differential form of this equation is obtained by eq.(\ref{orge2})
$$
\sum_{i} V_i(\bg)\frac{\partial \xi(\bg)}{\partial g_i} + \xi(\bg) =0, 
$$
which is well-known  as an equation for an invariant length scale 
$\xi(\bg)$.
Further, eq.(\ref{xi2}) is a natural generalization of the correlation 
length used in the 2D classical XY model.%\ref{BKT}.  

We consider the case where the running coupling constants are in the 
perturbative region $|\bg(t, \ba_0)| < 1$.  In this section, 
we study in particular the long-time asymptotics of a flow 
that once approaches the origin and then leaves 
for the non-perturbative region 
$|\bg(t, \ba_0)| \geq 1$, as the flow in the region $g_2 \geq |g_1|$ 
 in  the XY model. 
Generally,  a quadratic differential equation 
 such as eq.(\ref{orge2}) admits a flow 
qualitatively different from that investigated here.  
In section \ref{discussion}, we discuss 
such an exceptional case.

Next, let us represent $\xi(\ba_0)$ by the solution of the new 
RGE(\ref{rgrg}).  Define $\tau_{\rm R}$ by 
$e^{-\tau_{\rm R}} |\ba_0| = 1$.  
   From the definition of $s(\tau)$, we obtain 
\beq
        \log \xi(\ba_0) =  s(\tau_{\rm R}) 
        = \int_0^{\tau_{\rm R}} 
        d \tau \frac{d s}{d \tau}.  
\label{lxi}
\eeq
Using the differential equation (\ref{s}) on the right-hand side, 
we obtain the integral representation for $\xi(\ba_0)$. 
Since a flow treated in this section has a turning point 
as shown in Fig.\ref{global},  the correlation length is represented via
%Every turning point on the flow (\ref{inita}) corresponds 
%to a branch point of the mapped flow on $S$.  
%For example,  as in Fig. 1, 
%if $\bg(t, \ba_0)$ goes through one turning point $\bar{\bg}$, 
%the correlation length is represented by 
\beq
        \log \xi(\ba_0) = - \int_0^{\bar\tau} \, d\tau 
        \frac{e^\tau \, a^2}
        {\ba^{(1)}({\tau}) \cdot \bV\left(\ba^{(1)}({\tau})\right)}  
        - \int_{\bar \tau}^{\tau_{\rm R}} \, d\tau 
        \frac{e^\tau \, a^2}
        {\ba^{(2)}(\tau) \cdot \bV\left(\ba^{(2)}(\tau)\right)}.  
\label{int}
\eeq

Employing the integral representation, 
we argue that the leading term of $\xi$ is given by
\beq
        \log \xi(\ba_0) \simeq  e^{\bar{\tau}}
\label{mvf}
\eeq
if $\bar{\tau}$ in eq.(\ref{bartau}) is sufficiently large. 
Even though the integral near the turning point seems to 
diverge, it is only apparent as discussed in the previous section.  
\footnote{
In fact,  there exists a component of $\bV$ which 
does not vanish at the turning point, say,  $V_1$.  
We parametrize the flow of the new RGE by  $a_1$ instead of 
$\tau$ near the turning point and change the integration variable 
in eq. (\ref{int}).  
The measure changes as  
\beq
        \frac{d \tau}{\ba ({\tau}) \cdot \bV\left(\ba ({\tau})\right)} 
        = \frac{da_1}{a_1\ba \cdot \bV\left(\ba \right) - a_0^2 V_1(\ba)}  
\nn
\eeq
from eq. (\ref{rgrg}).  
The denominator of the right-hand side does not vanish near 
the turning point.  Therefore the denominators in the integrand in 
eq.(\ref{int}) does not contribute to the divergence of $\xi(\ba_0)$ 
and can be replaced by a certain constant when we evaluate the leading 
divergence of  
the integration in eq.(\ref{int}). 
}
The first term in the right-hand side  
of eq.(\ref{int}) diverges due to $e^\tau$ in the integrand 
when $\bar{\tau}$ goes to infinity. 
The second term contributes to the 
correlation length with the same order as the first term.
Hence, we can evaluate $\xi(\ba_0)$ by eq.(\ref{mvf}), which
translates singular behavior of $\xi(\ba_0)$ 
into that of $\bar{\tau}$.  
%Generalization of this argument to 
%the case when there are multiple turning points on 
%$\bg(t, \ba_0)$ is straightforward.  

Next, we evaluate the divergent $\bar{\tau}$ by using 
the polar-coordinate expression eq.(\ref{polar}) of the new RGE.
It is obvious that $\bar{\tau}$ diverges if 
the initial coupling constant $\ba_0$ is on an incoming straight flow line. 
It implies that $\bar\tau$ grows when $\ba(\tau)$ passes near 
a fixed point $\ba^*$ on the incoming straight flow line. 

\begin{figure}[ht]
\setlength{\unitlength}{1mm} 
        \begin{picture}(150, 65)(-30,-3)
        \put(-15,0){\epsfysize=6.0 cm \epsfbox{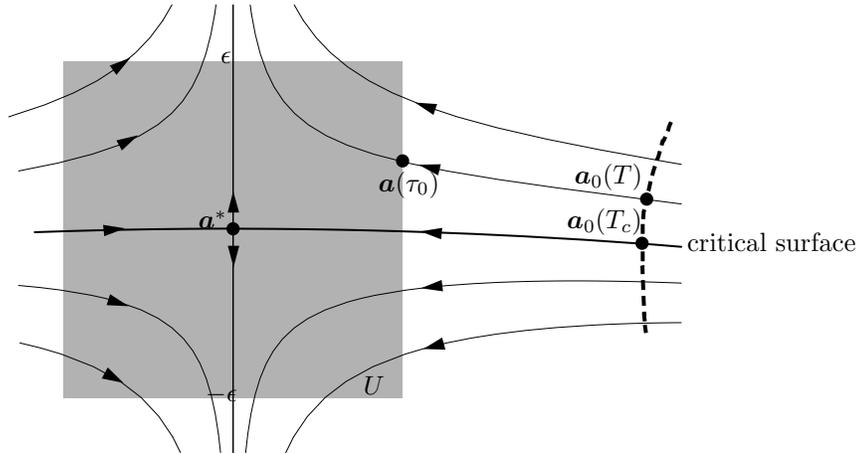}}
        \put(43,52){$\epsilon$}
        \put(41,7){$-\epsilon$}
        \put(64,35){$\ba(\tau_0)$}
        \put(40,30){$\ba^*$}
        \put(62,8){$U$}
        \put(89,30){$\ba_0(T_c)$}
        \put(90,36){$\ba_0(T)$}
                \put(105,27){critical surface}
        \end{picture}

\caption{Flow in the new RGE (\ref{rgrg}) near the fixed point $\ba^*$ 
on an incoming straight flow line. Dashed line represents 
one-parameter family of the initial value $\ba_0(T)$. 
}
\label{flowonS}
\end{figure}

Suppose that $\ba(\tau)$ goes through a neighborhood $U$ of $\ba^*$, 
as shown in Fig.\ref{flowonS}. 

%%%%%%%%%%%%%%%%%%%%%%%
% Fig: flowonS
%%%%%%%%%%%%%%%%%%%%
%%%%%%%%%%%%%%%%%%%%%%%
% End of flowonS
%%%%%%%%%%%%%%%%%%%%

The scaling matrix of the beta function (\ref{polar}) does not 
vanish at the fixed point in general and then the beta function 
can be linearized in $U$.  That is, 
\beqa
        \left( f_\al (\ba) \right)^{-1} 
        \tilde{\beta}_\al (\ba) &\simeq& \sum_{\ga=1}^{n-1} 
        \left( f_\al (\ba^*) \right)^{-1}
        \frac{\del \tilde{\beta}_\al}{\del \theta_\ga} (\ba^*)
        \delta \theta_\ga \nn\\
        &\equiv&
        \sum_{\ga=1}^{n-1} 
        A_{\al \ga} (\ba^*) \delta \theta_\ga,  
\label{scmt}
\eeqa 
where $\delta \theta_\ga \equiv \theta_\ga-\theta_\ga^* $ with 
$\{ \theta^*_\ga \}_\ga$ representing the fixed point $\ba^*$. 
The scaling matrix $A_{\al \ga}(\ba^*)$ describes the large 
$\bar{\tau}$ behavior because $\ba(\tau)$ spends a long time in $U$. 
If the scaling matrix  is diagonalized  with eigenvalues 
$b_\al$ by a new coordinate $\{ \theta'_\al \}_\al$, 
the new RGE becomes
\beq
\frac{d}{d \tau} \delta \theta'_\al = b_\al \delta \theta'_\al.
\label{lrgrg}
\eeq
The solution is  
\beq
        \delta \theta'_\al (\tau) 
        =   \delta \theta'_\al (\tau_0) e^{b_\al (\tau-\tau_0)}.  
\label{app} 
\eeq
 We take $U$ as the $(n-1)$-dimensional cubic box, whose side is 
$2 \epsilon$ and center is $\ba^*$. 
If the scaling matrix has the unique relevant mode $\theta'_1$, 
$\ba (\tau)$ spends time 
\beq
        \frac{1}{b_1} \log \left| \frac{\epsilon}{\de \theta'_1(\tau_0)}
        \right|  
\label{time}
\eeq
in $U$. Here we have supposed that $\ba (\tau)$ reaches 
$U$ at $\tau=\tau_0$.  
Now we vary the initial value $\ba_0$ by one parameter $T$ and 
assume that $\ba_0 (T)$  intersects a critical surface transversally 
at $T= T_c$.  See Fig.\ref{flowonS}.  

As the initial value $\ba_0(T)$ tends to the critical surface, 
$| \de \theta'_1(\tau_0)| $ gets small. 
%As the initial value $\delta \theta_\al (0)$ is closed 
%to a critical surface $\de \theta'_1 =0$,  
%$\de \theta'_1(\tau_0)$ tends to zero. 
It implies that $\de \theta'_1(\tau_0)$ is expanded as 
\beq
        \de \theta'_1(\tau_0) = {\rm const.} \left( T- T_c \right) + 
        {\rm O}\left(\left( T- T_c \right)^2 \right).  
\label{T-T_c}
\eeq
Since $\ba(\tau)$ spends a finite time outside of $U$, the divergent 
behavior of $\xi(\ba_0)$ is determined by eqs. (\ref{time}) and 
(\ref{T-T_c}).  Thus we get 
\beq
        \log \xi(\ba_0) \simeq e^{\bar{\tau}} \simeq 
        \left|T - T_c \right|^{1/b_1}, 
\eeq
which means 
\beq
        \sigma = \frac{1}{b_1}. 
\label{cef}
\eeq
This quantity does not depend strongly 
%either 
on the choice of $\ba_0(T)$,  
%or on the choice of the $n-1$ dimensional surface, 
so that $\sigma$ is a universal quantity 
in this sense. 

It is quite useful to find a relationship between the scaling matrix 
$A(\ba^*)$ in eq.(\ref{scmt}) and 
the $n \times n$ matrix 
\beq
         B_{i j} (\ba^*) \equiv \frac{\del \be_i }{\del a_j}  (\ba^*)
\label{B}
\eeq
for practically computing the eigenvalues $\{b_\al\}_\al$.  
In appendix \ref{AandB.app}, we will show that 
\beq
        \Lambda(B) = \Lambda(A) \cup \{ 0 \}, 
\label{AandB}
\eeq
where $\Lambda(M)$ is the set of eigenvalues of a matrix $M$. 
It should be noted that $B(\ba^*) =  B(-\ba^*)$ since $\bbe(\ba)$ is 
an odd function. This means that the scaling matrix $A(\ba^*)$ has the 
same eigenvalues as $A(-\ba^*)$. 

Now we deal with the 2D classical 
XY model again and show how to derive $\sigma$ by our method.  
The original RGE is given by eq.(\ref{xy}).  
We saw in the previous section that 
$\ba^* = (a_0/\sqrt{2}, a_0/\sqrt{2})$ is 
a fixed point on the incoming straight flow line.  
The matrix $B$ defined by eq.(\ref{B})  is 
\beq
        B(\ba^*) = \frac{\del \beta_i}{\del a_j} (\ba^*) = 
        \left(
        \begin{array}{rr}
                1 & -1 \\
                -1 & 1
        \end{array}
        \right) 
\sim  
        \left(
        \begin{array}{rr}
                0 & 0 \\
                0 & 2
        \end{array}
        \right),  
\label{matrixxy}
\eeq 
i.e., the scaling matrix $A(\ba^*)$ has the eigenvalue 2, 
according to eq.(\ref{AandB}).   
Alternatively, $A(\ba^*)$
is directly computed in this case.  From eq.(\ref{polarxy}) we get 
\beq
        A(\ba^*) = - \frac{1}{2} 
        \left. \frac{d }{d \theta} \right|_{\theta=\pi/4} 
         \cot 2\theta = 2,   
\eeq
as expected. 

Following our result eq.(\ref{cef}),  we get
\beq
        \sigma = 1/2, 
\eeq
which is well-known as the BKT universality.
  Originally it is obtained 
integrating the nonlinear RGE (\ref{xy}) explicitly\cite{KW}.   In contrast, 
according to our approach, we can reach the same result 
in an algebraic way.   As we 
will see in section \ref{examples},   it can provide $\sigma$
 even 
in the case when an original RGE(\ref{orge2}) is not integrable.

%If the fixed point  $\ba^*$ has the eigenvalues $b_1, b_2, \cdots $,
%the other fixed point $-\ba^*$ has the eigenvalues $-b_1, -b_2, \cdots $ 
%in the 
%RG transformation ${\cal R}_\tau$ defined by
%eq.(\ref{rgt}). If we have $m$ relevant operators and $n$ irrelevant 
%operators in the new RG
%near a fixed point $\ba^*$, we have $n$ relevant operators
%and $m$ irrelevant operators near the other fixed point $-\ba^*$.

%Note that the operator product coefficient $b_1 ^k$ affects
%the value of the critical exponet, eventhough it is 
%marginally irrelevant operator's. 

%Even in more general case of regions of the coupling parameter space,
%the critical exponent is determined 
%only by the operator product coefficients
%of two marginal operators.

In the above example,  the fixed point $\ba^*$ has a 
unique relevant mode,  so that we can apply the result (\ref{cef}).  
If there are multiple relevant modes in the scaling matrix, 
we can observe other relevant exponents 
$1/b_2, \ 1/b_3, \cdots > 0$ 
in an appropriate fine tuning of the initial parameters.

Finally,  we discuss the irrelevance of the higher-order terms in
the original RGE (\ref{orge2}). 
Here, we assume that we acquire no extra fixed points 
by taking into account higher-order terms.
If we have 
higher-order terms, the RG transformed coupling 
with $\tau$ obeys a different equation because of their scale breaking
nature. The scaled coupling $\bg'(t) = e^\tau \bg (e^{\tau} t, \ba_0)$
obeys 
\begin{equation}
\frac{d \bg'}{d t} = \bV(\bg') + {\rm O}(e^{-\tau}{\bg'}^3).
\end{equation}
Note that the higher-order term becomes smaller and
the RGE takes the scale invariant form asymptotically.
Therefore higher-order terms are irrelevant to determine the
critical exponent. 

\section{Logarithmic dependence of multiple 
marginally irrelevant coupling constants}
\label{massless}

So far, we have studied solutions
 of RGE (\ref{orge2}) in a massive phase. 
Our method is also applicable to studying asymptotic 
behavior of a solution in a massless phase.  
In this section, 
we study the logarithmic dependence of the multiple 
running coupling constants in a massless phase. 

It is important to clarify  
finite-size corrections in a system with marginally irrelevant 
 operators. For example, a numerical simulation in a spin system
 can calculate energy levels
 only for small degrees of freedom. 
 A theoretical formula for the finite-size
 correction is useful to extrapolate numerical data
 to those in the infinite system.
 If the system can be described in a critical theory with 
 marginally irrelevant perturbations, physical quantities 
 acquire logarithmic finite-size corrections.
 Here we are interested in a
system with a finite volume $L^D$ described by a theory 
with marginally irrelevant coupling constants $\bg$,
where $D$ is the space dimension.
Consider the situation where we have a critical theory with $\bg={\bf 0}$ 
which describes the system with an infinite volume.
In the system with a finite volume $L^D$, we can calculate 
physical quantities with a finite size correction
in the critical theory with a small 
perturbation of the  coupling $\bg$ 
obeying RGE (\ref{orge2}).
If we have an initial coupling $\ba_0$ at a lattice spacing 1,
the running coupling at the scale $L$ becomes $\bg(\log L, \ba_0)$, 
where $\bg(\infty, \ba_0)={\bf 0}$.
In the case of a single marginally irrelevant coupling constant $g$, 
the original beta function is given by 
$$
V(g)= C g^2 + {\rm O}(g^3),
$$ 
where $C$ is a universal constant in the sense that it is independent of
an initial value.  
The running coupling constant with an initial condition
$a_0$ has the following solution
\beq
\log L = \int_{a_0} ^{g(\log L, a_0)} \frac{dg}{V(g)} 
\simeq \frac{1}{C g(\log L,a_0)}-\frac{1}{C a_0}.
\eeq 
%where $L$ is the length scale of our interest, 
%has the following expansion: 
In this solution, we have a well-known universal expression \cite{C}
\beq
        g(\log L, a_0) = \frac{1}{C} \frac{1}{\log L} + 
        {\rm O}\left(\frac{1}{(\log L)^2},
        \frac{\log \log L}{(\log L)^2} \right). 
\label{single}
\eeq
The leading term is independent of the initial coupling $a_0$,
and therefore this formula is 
useful to fit numerical or experimental data of the system with a finite 
size.
For example, in one dimensional quantum spin systems
with marginally irrelevant perturbations,
logarithmic finite-size corrections to the ground state energy
$$
\Delta E_0 = - \frac{\pi}{6 L} \left( c + \frac{A}{(\log L)^3} +
{\rm O}\left(\frac{\log \log L}{(\log L)^4}, \frac{1}{(\log L)^4} \right) 
\right)
$$ 
are calculated from this formula (\ref{single}) \cite{C,AGSZ}, 
where $c$ is the central charge and $A$ is
determined from $C$. Since $c$ and $A$ are universal constants,
%A universal number such as $C$ can be 
%detected from numerical simulations or experiments.   
we can compare $c$ and $A$ to numerical (experimental)
data and obtain a clue whether 
a field theory that derives eq.(\ref{single}) 
is truly effective or not. 
Therefore it is important to derive a formula corresponding to 
eq.(\ref{single}) where there are multiple 
marginally irrelevant couplings. 
In this section,  we shall show that this universal nature holds in 
this case, as well.

As we mentioned above, 
we examine the case where the all coupling constants 
are marginally irrelevant, so that  flow of eq.(\ref{orge2}) 
is absorbed into the origin.   In this case, there are 
no turning points on the flow,  which implies that 
the transformation ${\cal R}_\tau \ba_0$ defined in eq.(\ref{rgt}) is 
single-valued with respect to $\tau$.  Therefore we can write 
down a formula similar to eq.(\ref{int}) as 
\beq
        \log L = - \int_0^\tau d \tau'
        \frac{a_0^2 \, e^{\tau'}}{\ba(\tau') \cdot \bV(\ba(\tau'))}. 
\label{logL}
\eeq
The running coupling constant $\bg(\log L, \ba_0)$ is obtained by 
\beq
         \bg(\log L, \ba_0) =  e^{-\tau} \ba(\tau).  
\label{atog}
\eeq
In order to derive the logarithmic dependence of $\bg(\log L, \ba_0)$, 
we first solve eq.(\ref{logL}) for $\tau$ when $L$ is sufficiently large. 
Then we apply the result to eq.(\ref{atog}). 

As we have seen in the previous section, 
when we take $L$ sufficiently large, 
the contribution from a neighborhood $U$ of a fixed point 
$\ba^*$ on an incoming straight flow line 
dominates in the integration of eq.(\ref{logL}), which can be 
evaluated from the linearized new RGE in $U$.  
Suppose that $\ba(\tau)$ enters $U$ 
at $\tau=\tau_0$. 
Eq.(\ref{logL}) becomes 
\beq
        \log L \simeq - \int_{\tau_0}^\tau d \tau'
        \frac{a_0^2 \, e^{\tau'}}{\ba(\tau') \cdot \bV(\ba(\tau'))} 
\label{int2}
\eeq
for large $L$. 

Writing  
\beq
        \ba(\tau) = \ba^* + \de \ba(\tau),
\label{deba}
\eeq 
$\de \ba(\tau)$ in the polar-coordinate representation 
obeys the linearized RGE (\ref{lrgrg}) in $U$.
Its solution has the following asymptotic form for large $\tau$
\beq
\de \ba(\tau) \simeq \sum_{\al = 1}^{n-1}
\frac{\del \ba}{\del \theta'_\al} \de \theta'_{\al}(\tau)
\simeq  \frac{\del \ba}{\del \theta'_1} \de \theta'_{1}(\tau_0) 
e ^{b_1 (\tau-\tau_0)},
\label{app3}
\eeq
where $b_1<0$ is the maximal eigenvalue
of the scaling matrix $A(\ba^*)$ defined in eq.(\ref{scmt}).
We expand the integrand in eq.(\ref{int2}) as 
\beq
        \frac{a_0^2 \, e^{\tau'}}{\ba(\tau) \cdot \bV(\ba(\tau))} = 
        \frac{a_0^2 \, e^{\tau'}}{\ba^* \cdot \bV(\ba^*)}  
        \left(1 - \frac{(\de \ba \cdot \left. \nabla 
        \right|_{\ba = \ba^*})\ba^* \cdot \bV(\ba)}{\ba^* \cdot 
        \bV\left(\ba^*\right)} + {\rm O}\left(\de \ba^2 \right) \right), 
\label{app4}
\eeq
and calculate the right hand side of (\ref{int2}).
First we  compute the leading-order contribution. 
The leading integration is easily performed as follows:
\beqa
        \log L \simeq 
        - \int_{\tau_0}^\tau d\tau'
        \frac{a_0^2 \, e^{\tau'}}{\ba^* \cdot \bV(\ba^*)} &=& 
        -\frac{a_0^2}{\ba^* \cdot \bV(\ba^*)}(e^\tau -e^{\tau_0} )
        \simeq -\frac{a_0^2}{\ba^* \cdot \bV(\ba^*)} e^\tau. 
\label{leading}
\eeqa
Since  
$\ba \cdot \bV(\ba)$, which is a cubic function of $\{a_k \}$, 
 is negative at $\ba^*$, 
we can write  
\beq
        \ba^* \cdot \bV(\ba^*) =  - C a_0^3, 
\label{denomi}
\eeq 
where $C$ is a positive constant defined by 
\beq
        C \equiv        - \bee^* \cdot \bV(\bee^*)
\eeq
with $\bee^* \equiv \ba^*/a_0$.  
    From eqs. (\ref{atog}),
(\ref{leading}) and (\ref{denomi}), we get, in the leading order, 
\beq
        \bg(\log L,  \ba_0) = e^{-\tau} \ba^* \simeq \frac{1}{C \log L} 
        \bee^*.  
\eeq
Since  $\bee^*$ and $C$ are  completely determined by 
the explicit form of $\bV$,  the result in the leading order is universal.

Next, let us go to the next-to-leading term. After evaluating 
the next-to-leading term in the integral (\ref{int2})
with the help of eqs.(\ref{app3}) and (\ref{app4}), 
we represent $e^\tau$ in $1/\log L$ expansion.
%The result is 
%\beq
%        = \left \{
%       \begin{array}{ll}       
%       \left(1 - q  \frac{\log \log L}{\log L} + \cdots \right) 
%       \left( + c_1 \bee_1 \left(\log L \right)^{-1} +  
%       \cdots \right)&(b_1 = -1) \\ 
%    \frac{1}{C \log L }
%       \left(1 + c_0  \frac{1}{\log L} + \cdots \right) 
%       \left(\ba^* / a + c_1 \bee_1 \left(\log L \right)^{b_1} +  
%       \cdots \right)&(b_1 <-1) 
%       \end{array}
%       \right. , 
%\label{crr2}
%\eeq 
The calculation is easily performed and
finally we obtain 
\beq
\bg(\log L, \ba_0)=     \left \{
         \begin{array}{ll}
         \frac{1}{C \log L }\bee^* + 
         \frac{B }{\left(\log L \right)^{1-b_1}}
                \frac{\del \ba}{\del \theta'_1}& 
        (-1< b_1 < 0 ) \\
         \frac{1}{C \log L }\bee^* + 
         \frac{B' \log \log L}{(\log L)^2}
                \frac{\del \ba}{\del \theta'_1}& (b_1=-1)\\
        \frac{1}{C \log L }\bee^* + 
        \frac{B'' }{\left(\log L \right)^2}\bee^* & (b_1< -1)
                \end{array}
        \right.,  
\label{rslt4}
\eeq
where the constants $B$, $B'$ and  $B''$ depend on the initial 
condition, in contrast to the leading term. 
 The result implies that, if $-1 \leq b_1 < 0$, 
we have to take into account the non-universal nature  
of the subleading correction even though ${\rm O}(g^3)$ 
corrections in the original renormalization-group equations 
give universal coefficients to this subleading term. 
   
\section{Examples}
\label{examples}
 Here, we consider the level 1 $SU(N)$ Wess-Zumino-Witten (WZW) 
model in two dimensions as a critical theory\cite{BZ}. 
This model has traceless chiral currents 
$J^{a b}(z)$ and $\bar{J}^{ab}(\bar{z})$ $(a, b = 1, \cdots, N)$ satisfying
 the following operator product expansion (OPE): 
\beqa
        J^{a b}(z) J^{c d}(0) &\sim& \frac{\de_{a d}\de_{b c}}{z^2} + 
        \frac{1}{z} \left(\de_{b c} J^{a d}(z) - \de_{a d} J^{b c}(z)\right) \nn\\
        \bar{J}^{a b}(\bar{z}) J^{c d}(0) &\sim& 
        \frac{\de_{a d}\de_{b c}}{\bar{z}^2} + 
        \frac{1}{\bar{z}} \left(\de_{b c} \bar{J}^{a d}(\bar{z}) - 
        \de_{a d} \bar{J}^{b c}(\bar{z})\right) \nn\\
        J^{a b}(z) \bar{J}^{c d}(0) &\sim& 0, 
\label{kmalg}
\eeqa
where $\sim$ stands for equality up to regular terms. 
Using these currents, 
we can construct $(N^2-1)^2$ marginal operators 
$J^{ab}(z) \bar{J}^{cd}(\bar{z})$.  In this section,  we 
study models perturbed by some of the marginal operators which are 
inspired by a quantum spin chain\cite{AGSZ,IK}. 
\subsection{two-parameter system}
First, we consider a simple two-parameter system which includes the
BKT universality as the special case $N=2$.
We define the $SU(N)$ symmetric marginal operator $\phi^1(z, \bar{z})$
and the symmetry breaking one $\phi^2(z, \bar{z})$
\begin{equation}
\phi^1(z, \bar{z}) = \sum_{a=1} ^N \sum_{b=1}^N J^{ab}(z) 
\bar{J}^{ba} (\bar{z}), \ \ \ \ \ \ \ \  
\phi^2(z, \bar{z}) = \sum_{a=1} ^N \sum_{b=1}^N 
J^{ab}(z) \bar{J}^{ab} (\bar{z}),  
\end{equation}
which satisfy the following closed OPE
\begin{eqnarray}
\phi^1(z, \bar{z}) \phi^1(0, 0) &\sim& 
\frac{-2N}{|z|^2} \phi^1(0, 0) \nn\\ 
\phi^1(z, \bar{z}) \phi^2(0, 0) &\sim& \frac{-2}{|z|^2}\left( \phi^1(0, 0) 
- \phi^2(0, 0) \right) \nn\\ 
\phi^2(z, \bar{z}) \phi^2(0, 0) &\sim& 
\frac{2N}{|z|^2} \phi^2(0, 0) 
\label{OPE}
\end{eqnarray}
by eq.(\ref{kmalg}). 
The action integral $\cal{A}$ of the perturbed theory is
\begin{equation}
{\cal A} = {\cal A}_{\rm WZW} + \sum_{i=1} ^2 
g_i \int\frac{d^2z}{2 \pi} \phi^i(z, \bar{z}).
\end{equation}
The OPE formula eq.(\ref{OPE})
yields the following two-parameter RGE: 
%\begin{equation}
% \left( \begin{array}{cc} 
%b_1^1 & b_1^2 \\
%b_2^2 & b_2^2 
%\end{array} \right) = 
%\left( \begin{array}{cc}
%1  &  -2/N \\
%-2/N &  1
%\end{array} \right).
%\end{equation}
\begin{eqnarray}
\frac{d g_1}{d t} &=& V_1(\bg) = g_1 (N g_1 + 2 g_2) \nonumber\\ 
\frac{d g_2}{d t} &=& V_2(\bg) = - g_2 (2 g_1 + N g_2). 
\label{2prmex}
\end{eqnarray}
In the case of $N=2$, 
the RGE reduces to the same form as the RGE of the XY model with an 
appropriate linear transformation, which was extensively studied 
in sections \ref{rgeforrge} and \ref{massive}.  
Here we restrict ourselves to the case of $N \geq 3$.   
 
The beta function  in the new RGE  (\ref{rgrg}) for eq.(\ref{2prmex}) 
is 
\beqa
        \beta_1(\ba) &=& a_1 -\frac{a_1(N a_1 + 2 a_2) a_0^2
}{ a_1 ^2(N a_1+2 a_2) - a_2 ^2(2 a_1 +N a_2)}, \nonumber \\
\beta_2(\ba) &=& a_2 -\frac{-a_2(2 a_1 + N a_2) a_0^2
}{ a_1 ^2(N a_1+2 a_2) - a_2 ^2(2 a_1 +N a_2)}.
\eeqa
  Solving $\bbe(\ba) ={\bf 0}$, we have the following six fixed points: 
$\pm(a_0, 0), \ \pm(0, a_0)$,  and  $\pm(a_0/\sqrt{2}, -a_0/\sqrt{2})$. 
Evaluating $\ba \cdot V(\ba)$ at those points, we find that 
there are the three fixed points on incoming straight flow lines, 
$(-a_0, 0)\equiv {\bf c}_1, \ (0, a_0)\equiv {\bf c}_2$,  
and  $(-a_0/\sqrt{2}, a_0/\sqrt{2})\equiv {\bf c}_3$. 
The matrix $B(\ba)$ in eq.(\ref{B}) at those points becomes 
\beq
        B({\bf c}_1) =  
        \left(
        \begin{array}{cc}
                0 & 0 \\
                0 & \frac{2+N}{N}
        \end{array}
        \right), \  
                B({\bf c}_2) =  
        \left(
        \begin{array}{cc}
                 \frac{2+N}{N}& 0\\
               0 & 0
        \end{array}
        \right), \  
        B({\bf c}_3) =  \frac{2+N}{4-2N}
        \left(
        \begin{array}{cc}
                1 & 1\\
                1 & 1
        \end{array}
        \right) 
\sim  
        \left(
        \begin{array}{cc}
                0 & 0 \\
                0 & \frac{2+N}{2-N}
        \end{array}
        \right),  
\label{egnvl}
\eeq 
which means  ${\bf c}_1$ and 
${\bf c}_2$ are unstable fixed points while 
${\bf c}_3$ is stable for all $N \geq  3$. 

Divergence of 
the correlation length is governed by the unstable fixed 
points and, according to the formula eq.(\ref{cef}),  
\begin{equation}
\sigma = \frac{N}{N+2}, 
\label{sigma}
\end{equation}
which is identical to that obtained by the explicit solution 
of the differential equation in Ref.\cite{IK}.  Note that the result 
eq.(\ref{sigma}) is also valid in the case of $N=2$ although 
the scaling matrix cannot be defined at ${\bf c}_3$: ${\bf c}_3$ 
corresponds to a fixed line of the original RGE (\ref{2prmex}) 
if $N=2$.  

Any theory of $c=1$ CFT with marginal perturbations 
has the critical exponent $\sigma=1/2$ or $\sigma=1$, 
as is well-known \cite{Kd}.
This is because the level 1 $SU(2)$ WZW theory is the
maximally symmetric theory in $c=1$ CFT and because 
it gives the most general
theory with marginal perturbation in $c=1$ CFT \cite{G}. The most general
theory with marginal perturbations describes a quantum XYZ chain
with spin 1/2. 
The infinite-order phase transition occurs at a line of the
XXX chain. This corresponds to $N=2$ in our analysis.  
In the case of $c > 1$ CFT with marginal perturbation, however,
we show some new universality classes
with non-trivial critical exponents 
$\sigma$ eq.(\ref{sigma}) for $N > 2$.
For example, the transition in the case of $N=3$ describes
the gapless Haldane gap phase transition with the exponent 
$\sigma = 3/5$ from the $SU(3)$ symmetric line $g_2=0$ 
in an isotropic spin 1 chain \cite{IK}.

\begin{figure}[ht]
\setlength{\unitlength}{1mm}
 \begin{picture}(150, 60)(-15,0)
        \put(-5,0){\epsfxsize=7cm \epsfbox{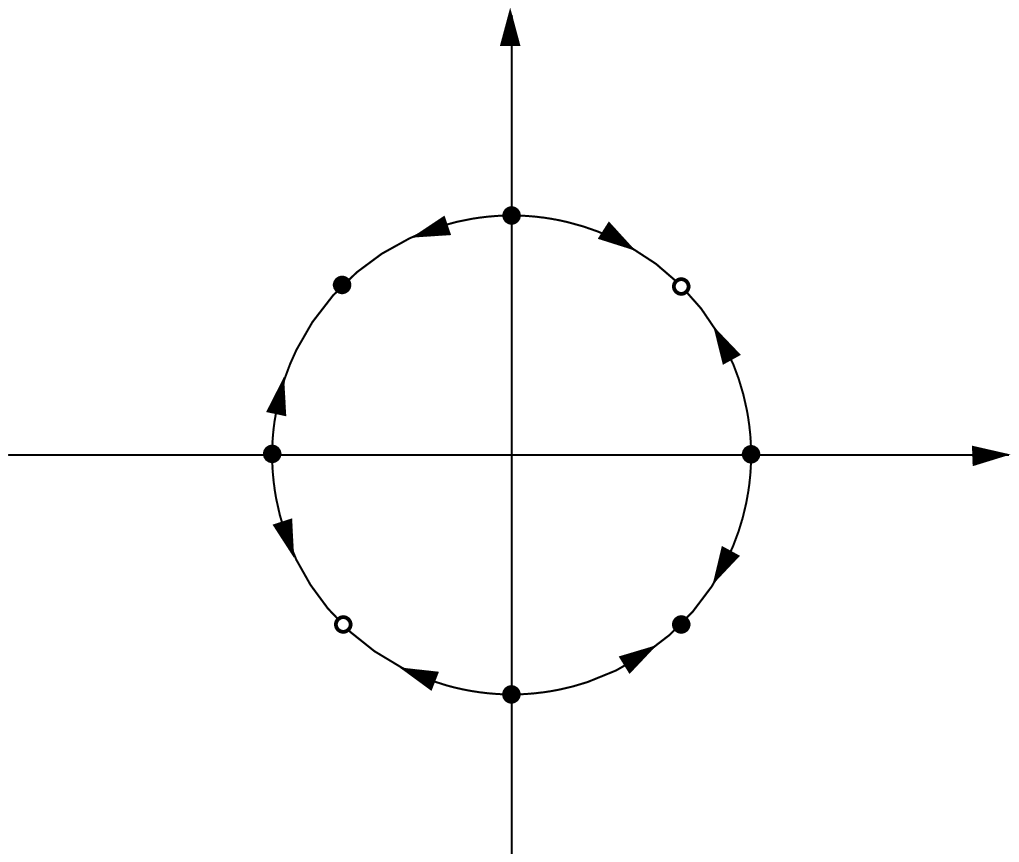}}
                \put(75,0){\epsfxsize=7cm \epsfbox{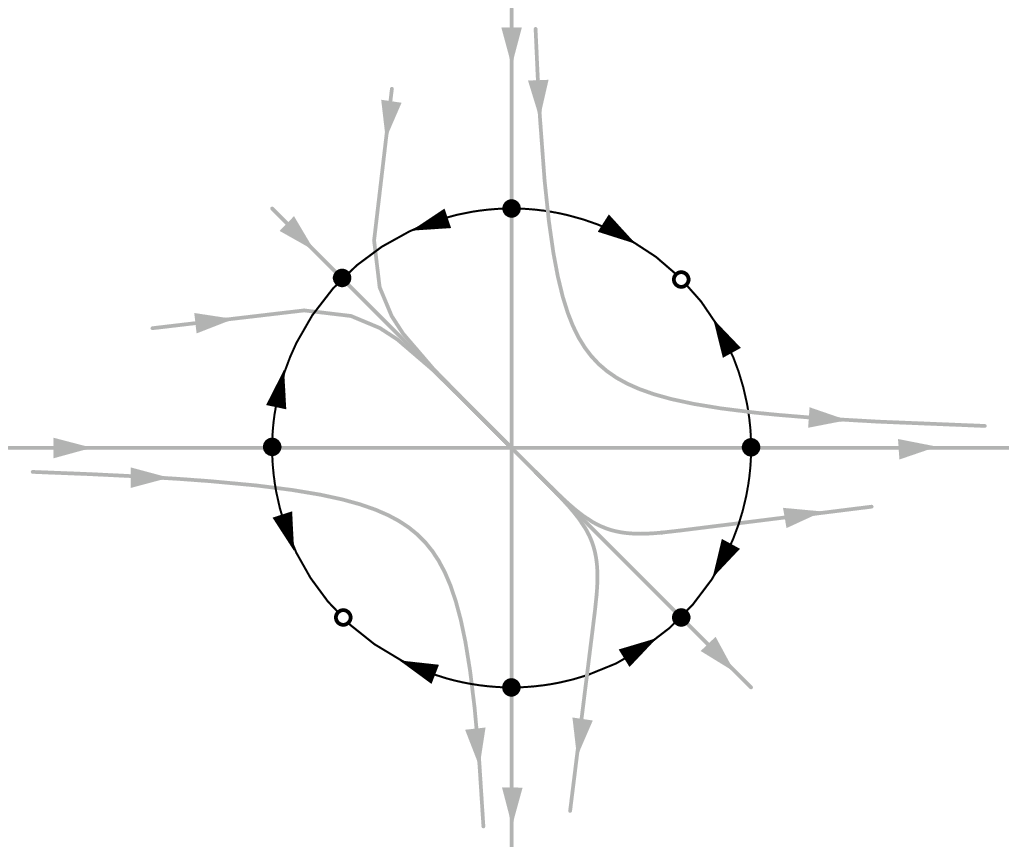}}
                \put(65,27){$g_1$}
        \put(31,60){$g_2$}
        \put(27,46){${\bf c}_1$}
        \put(107,46){${\bf c}_1$}
        \put(10,29){${\bf c}_2$}
        \put(90,29){${\bf c}_2$}
        \put(14,40){${\bf c}_3$}
        \put(94,40){${\bf c}_3$}
                \put(52,47){(a)}
                \put(132,47){(b)}
\end{picture}
\caption{(a) Flow in the new RGE.  The black circles stand for 
fixed points while the white ones branch points corresponding to 
turning points on the flow in eq.(\ref{2prmex}). (b) Illustration
 of the flow in 
eq.(\ref{2prmex}) derived by the new RGE, which is 
drawn in gray curves.} 
\label{2prmflw}
\end{figure}

The result eq.(\ref{egnvl}) is also useful when we  figure out 
the qualitative picture of the flow in the original RGE (\ref{2prmex}). 
For this purpose, we need to know a branch point, which corresponds 
to a turning point on a flow in eq.(\ref{2prmex}), by solving  
$\ba \cdot \bV(\ba)=0$.   The solution is 
 $\pm (a_0/\sqrt{2}, a_0/\sqrt{2})$ for all $N \geq 3$.  A 
 flow in the new RGE changes its direction 
at these points, as we depicted 
in Fig.\ref{global}.  Combining the result eq.(\ref{egnvl}) and 
the fact that
the scaling matrix at $-{\bf c}_i$ $(i=1,2,3)$ has the same eigenvalues 
as at ${\bf c}_i$, we
get the global  flow of the new RGE,  as in Fig.\ref{2prmflw}(a).  
It derives qualitative features of the RG flow in eq.(\ref{2prmex}), 
which are drawn in gray curves in Fig.\ref{2prmflw}(b). 
We notice that the region $g_1 < 0$, $g_2 >$ is a massless phase, 
where solutions in the original RGE are all absorbed into the origin along 
the incoming straight flow line $g_1 + g_2 =0$ $(g_1 < 0)$. 
The incoming straight flow lines passing ${\bf c}_1$ or ${\bf c}_2$ 
form the phase boundary. 
%%%%%%%%%%%%%% 
%   Fig: 2prmflw
%%%%%%%%%%%%%%

Next, we discuss  logarithmic dependence 
of the running coupling constants in the massless phase. 
Let us introduce new variables 
$(X, \, Y) \equiv (g_1-g_2, \, -g_1 - g_2 )$.  
The original RGE has the incoming straight flow line 
$Y=0$ $(X<0)$ on which ${\bf c}_3$ is situated. According to our result 
eq.(\ref{rslt4}),  the running coupling constant $X(\log L, \ba_0)$ 
has the leading logarithmic 
dependence $1/\log L$, whose coefficient is universal. In contrast, 
$Y(\log L, \ba_0)$ has the dependence $(\log L)^{-1+b}$ with a non-universal 
coefficient, where $b\equiv (N+2)/(2-N) < -1$ for $N \geq 3$. Hence 
the $1/(\log L)^2$ contribution that belongs the next-to-leading term 
in $X (\log L, \ba_0)$ 
gives sub-leading contribution.  This implies that,  
if we can determine $g^3$ terms in the original 
RGE (\ref{2prmex}), universal $\log \log L/(\log L)^2$ dependences can be 
obtained in this example.  We remark that the logarithmic dependence of 
$Y(\log L, \ba_0)$ is consistent with the result from the explicit 
solution \cite{IK}. 

%$g_1 < 0, g_2 > 0$. The original RGE has the incoming  straight flow line 
%$Y \equiv -g_1 - g_2 =0$ for $N \geq 3$. 
%The coupling constant $X \equiv g_1-g_2$
%has the leading logarithmic dependence 
%$1/\log L$.
% and  $Y$ has sub-leading term 
%$\left( 1/\log L \right)^{1-b_1}$.
%The eigenvalue of the beta function (\ref{rgrgex1})
%at the fixed point $(1/\sqrt{2}, -1/\sqrt{2})$
%corresponding to $Y=0$ straight flow line
%is $b_1 = (2+N)/(2-N)$ which agrees with 
%that obtained by the exact solution \cite{IK}.
%Since $b_1= (2+N)/(2-N) < -1$, $1/(\log L)^2$ term gives
%sub-leading contribution in this case. 

\subsection{three-parameter system}
Here, we consider a non-trivial three-parameter system, 
an $SU(2)$-invariant marginal perturbation 
of the level 1 $SU(4)$ WZW model whose RGE becomes nonintegrable.  
This model may describe an $S=3/2$ quantum spin chain around 
the $SU(4)$ symmetric 
Uimin-Lai-Sutherland model \cite{ULS,PT} with some $SU(2)$ invariant 
perturbation.  The $SU(2)$ transformation 
is generated by 
\beq
        {\rm Tr} \int d z  J(z)  \bL  + {\rm Tr} \int d \bar{z} 
\bar{J}(\bar{z}) \bL, 
\eeq
where $\bL = (L^1, L^2, L^3)$  is the spin matrix in the spin-3/2 
representation.  Marginal operators invariant under the 
$SU(2)$ transformation is constructed as follows:  
\beq
        \phi^j(z, \bar{z}) \equiv  
        \Tr  \sum_{m=-j}^{j} (-1)^m J(z) T_{j,m} \bar{J}(\bar{z}) T_{j,-m}, 
        \qquad j = 0, 1, 2, 3, 
\eeq
where $T_{j,m}$ satisfies 
$[\bL^2, T_{j,m}] = j(j+1) T_{j,m}$ and $[L^3, T_{j,m}] = m T_{j,m}$.  
Using the tracelessness property of the currents, we get  
\beq
        \sum_{j=0}^3 \phi^j(z, \bar{z}) = 0, 
\eeq
which indicates that there are three independent marginal operators in 
$\phi^0, \cdots, \phi^3$. 
Here we consider the perturbation 
\beq
        \sum_{j=0}^2 g_i \int \frac{d^2 z }{2 \pi} \phi^j(z, \bar{z}).  
\eeq
Employing the OPE (\ref{kmalg}) and the normalization condition 
$\Tr \, {}^t T_{j,m} T_{j', m'} = \de_{j j'} \de_{m m'}$,  we 
find the following operator product expansions
\beqa
        \phi^{0}(z,\bar{z}) \phi^{0}(0,0) &\sim& \frac{2}{|z|^2}
         \phi^{0}(0,0) 
        \nn\\
        \phi^{0}(z,\bar{z}) \phi^{1}(0,0) &\sim& \frac{1}{|z|^2}
        (-\frac{3}{5} \phi^{0}(0,0) - \frac{1}{2}\phi^{1}(0,0) - 
        \frac{3}{10}\phi^{2}(0,0)) 
        \nn\\
        \phi^{0}(z,\bar{z}) \phi^{2}(0,0) &\sim& \frac{1}{|z|^2}
        (-\frac{1}{2} \phi^{0}(0,0) + \frac{1}{2}\phi^{2}(0,0)) 
        \nn\\
        \phi^{1}(z,\bar{z}) \phi^{1}(0,0) &\sim& \frac{1}{|z|^2}
        (-\frac{6}{25} \phi^{0}(0,0) + \frac{11}{5}\phi^{1}(0,0) + 
        \frac{3}{5}\phi^{2}(0,0)) 
        \nn\\
        \phi^{1}(z,\bar{z}) \phi^{2}(0,0) &\sim& \frac{1}{|z|^2}
        (-\frac{9}{10} \phi^{0}(0,0) -\frac{1}{2}\phi^{1}(0,0) - 
        \frac{6}{5}\phi^{2}(0,0)) 
        \nn\\
        \phi^{2}(z,\bar{z}) \phi^{2}(0,0) &\sim& \frac{1}{|z|^2}
        (3 \phi^{0}(0,0) + 3 \phi^{2}(0,0)).  
\eeqa
The RGE derived from the OPE is
\beqa
 \frac{d g_0 }{dt} &=&  V_0(\bg) =
  -{{g_0}^2} + {{3\,g_0\,g_1}\over 5} + 
   {{3\,{{g_1}^2}}\over {25}} + \frac{g_0\,g_2}{2} + 
   {{9\,g_1\,g_2}\over 10} - \frac{3\,{{g_2}^2}}{2}
\nn\\
\frac{d g_1}{dt} &=&  V_1(\bg) =
  \frac{g_0\,g_1}{2}  - {{11\,{{g_1}^2}}\over 10} + 
   \frac{g_1\,g_2}{2}
\nn\\
\frac{d g_2 }{dt} &=& V_2(\bg) =
  {{3\,g_0\,g_1}\over 10} - {{3\,{{g_1}^2}}\over 10} - 
   \frac{g_0\,g_2}{2} + {{6\,g_1\,g_2}\over 5} - \frac{3\,{{g_2}^2}}{2}.  
\label{orgex3}
\eeqa
We find that the new RGE for eq.(\ref{orgex3}) have 
the following seven fixed points on incoming straight flow lines:
%\begin{eqnarray}
%\beta_0(\ba)&=&x-\frac{-2x(x-2z-6y)}{
%-2x^2(x-2z-6y)+y^2(x+2y-2z)
%+z(3z^2+24z y-6y x-z x)} \nonumber \\
%\beta_1(\ba)&=&y-\frac{y(x+2y-2z)}{
%-2x^2(x-2z-6y)+y^2(x+2y-2z)
%+z(3z^2+24z y-6y x-z x)} \nonumber \\
%\beta_2(\ba)&=&z-\frac{3z^2+24z y-6y x-z x}{
%-2x^2(x-2z-6y)+y^2(x+2y-2z)
%+z(3z^2+24z y-6y x-z x)}
%\beta_0(\ba) &=& a_0-
%\frac{V_0(\ba)}{a_0 V_0(\ba)+ a_1 V_1(\ba)+a_2 V_2(\ba)} 
%\nonumber \\
%\beta_1(\ba) &=& a_1-
%\frac{V_1(\ba)}{a_0 V_0(\ba)+ a_1 V_1(\ba)+a_2 V_2(\ba)} 
%\nonumber \\
%\beta_2(\ba) &=& a_2 -
%\frac{V_2(\ba)}{a_0 V_0(\ba)+ a_1 V_1(\ba)+a_2 V_2(\ba)} 
%\end{eqnarray}
%Zeros of the beta function 
\begin{eqnarray}
& & {\bf c}_1=(a_0, 0, 0),  \ 
{\bf c}_2=\left( \frac{a_0}{\sqrt{2}},0, \frac{a_0}{\sqrt{2}} \right), \
{\bf c}_3=
        \left(\frac{3a_0}{\sqrt{10}}, 0, \frac{a_0}{\sqrt{10}} \right), \
{\bf c}_4=\left( 3a_0 \sqrt{\frac{2}{35}}, a_0 \sqrt{\frac{5}{14}}, 
\frac{3a_0}{\sqrt{70}} \right), \nonumber \\ 
& &{\bf c}_5=\left(\frac{-3a_0}{5\sqrt{26}}, \frac{5a_0}{\sqrt{26}}, 
\frac{2a_0}{5}{\sqrt{\frac{2}{13}}} \right),\nonumber\\
&&{\bf c}_{\pm}= \frac{
a_0 \sqrt{25105 \mp 1747 \sqrt{205}}}{\sqrt{5678}}
\left(\frac{(\pm 239 \sqrt{5} +  85 \sqrt{41})}{180},
\frac{(\pm 13 \sqrt{5} + 5 \sqrt{41})}{12},  
\mp  \frac{1}{2 \sqrt{5}} \right). \nonumber
 \end{eqnarray}
%
%The minus sign to frac{1}{2 \sqrt{5}}is appended on 4/07/98
%
%
%
%The scaling matrix is defined by the derivatives of the beta functions
%at the fixt point. Note that the parameters $x, y, z$ are not independent 
%of each others because of the constraint $x^2+ y^2 +z^2 = 1$,
%and therefore we have two 
The corresponding eigenvalues of the scaling matrix 
at the each fixed points are calculated as
\begin{eqnarray}
{\bf c}_1;  && \left( \frac{3}{2}, \frac{1}{2} \right), \ \ \
{\bf c}_2;  \ \ \ \left( \frac{3}{2}, \frac{1}{2} \right), \ \ \
{\bf c}_3;  \ \ \ \left( \frac{5}{3}, \frac{-1}{3} \right), \ \ \
{\bf c}_4;  \ \ \ (-5, -11), \nonumber \\ 
{\bf c}_5;  && \left( \frac{55}{27}, \frac{41}{27} \right), \ \ \ 
{\bf c}_\pm; \ \ \ \left( \frac{-4+\sqrt{631}}{8},
\frac{-4-\sqrt{631}}{8} \right)
=(2.63996417 \cdots, -3.63996417 \cdots). \nonumber
  \end{eqnarray}
Namely, they are classified into the twice unstable fixed points 
$ {\bf c}_1,  {\bf c}_2,  {\bf c}_5$,
the once unstable fixed points  $ {\bf c}_3,  {\bf c_\pm}$
and the stable fixed point ${\bf c}_4$. 
%We have another set of six fixed points $-{\bf c}_i \ \ 
%(i=1, 2, \cdots, 5, \pm)$ which are out going 
%straight flow lines and have the same absolute value of 
%their cprresponding eigenvalues with opposite sign.
%Therefore, fourteen fixed points are classified into
%four twice unstable, six once unstable,
%and four stable  fixed points. 
The critical exponents determined by 
the once unstable fixed points ${\bf c}_3$ and ${\bf c}_\pm$ 
are respectively 
\begin{equation}
\sigma= \frac{3}{5}, \ \ \ \ \  \frac{8}{(-4+\sqrt{631})},   
\label{expnt}
\end{equation}
which must be observed most likely in this model.
%If the initial value crosses the critical surface,
%the coupling parameter on the sphere $S$
%running along the critical curve toward the once 
%unstable fixed point. The running coupling spends 
%the time almost in the critical region 
%near the once unstable fixed point and leaves away from
%the fixed point with the exponent (\ref{expnt}).
%The irrational value of the critical exponent 
%$\sigma= 8/(-4+\sqrt{631})$ is quite curiouse.
In order to detect other exponents, fine tuning of the initial 
coupling parameters
to the twice unstable fixed points    
is necessary as in the multi-critical behavior 
of ordinary second-order phase transitions.  

Logarithmic dependence of the running coupling constants is  
controlled by the stable fixed point $\bc_4$. Since 
the largest eigenvalue at the fixed point is less than $-1$, 
the sub-leading contribution is $1/(\log L)^2$ in this case. 

\begin{figure}[ht]
\setlength{\unitlength}{1mm}
 \begin{picture}(150, 35)(-15,0)
        \put(30,5){\epsfxsize=8cm \epsfbox{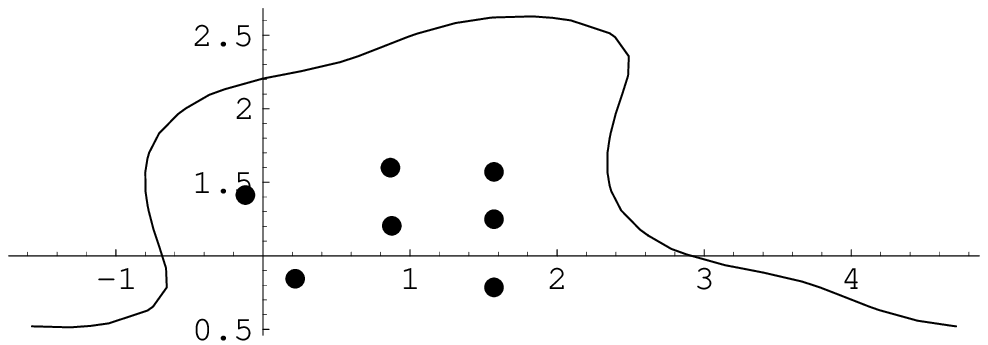}}
        \put(110,12){$\phi$}
        \put(51,33){$\theta$}
        \put(71,19){${\bf c}_1$}
        \put(71,15.5){${\bf c}_3$}
        \put(71,9){${\bf c}_2$}
        \put(62,20){${\bf c}_+$}
        \put(63,15){${\bf c}_4$}
        \put(52,17){${\bf c}_5$}
        \put(55,10){${\bf c}_-$}
\end{picture}
\caption{Seven fixed points ${\bf c}_1, \cdots ,{\bf c}_{\pm}$
}
\label{fxdpts}
\end{figure}

It is helpful to study the new RGE for eq.(\ref{orgex3}) numerically
 in order to better understand our results.
Fig.\ref{fxdpts} exhibits the seven fixed points
${\bf c}_1, \cdots ,{\bf c}_{\pm}$ in polar coordinates  
\beq
        \ba = a_0 \left(\sin \theta \, \sin \phi, \, \, \sin \theta \, 
        \cos \phi, \, \,  
        \cos \theta \right)
\eeq
with $-\pi/2 \leq \phi < 3 \pi /2$, $0 \leq \theta < \pi$. 
%%%%%%%%%%%%%% 
%   Fig: fxdpts
%%%%%%%%%%%%%%
%%%%%%%%%%%%%end of Fig:fxdpts%%%%%%%%%%%%

 The curve 
in Fig.\ref{fxdpts} represents the equation $\ba \cdot \bV(\ba) =0$,  
which corresponds to  the set of branch points of 
the RG transformation ${\cal R}_\tau$.   
All fixed points belong to the region 
$\{\ba \, | \, \ba \cdot \bV(\ba) <0 \}$ 
because they are on incoming straight flow lines in the original RGE. 

\begin{figure}[ht]
\setlength{\unitlength}{1mm}
 \begin{picture}(150, 55)(-15,-2)
        \put(30,0){\epsfxsize=8cm \epsfbox{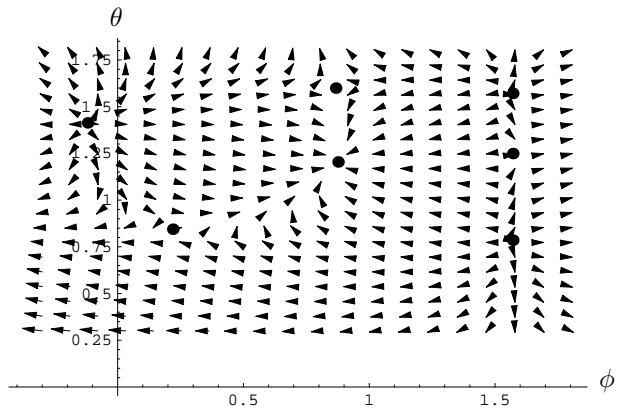}}
                \put(110,3){$\phi$}
                \put(45,51){$\theta$}
\end{picture}
\caption{The vector field defined by the polar-coordinate representation 
of the new RGE for eq.(\ref{orgex3}). The black dots stand for the fixed points, 
which correspond to those in Fig.\ref{fxdpts}. 
}
\label{vctfld}
\end{figure}

Fig.\ref{vctfld} shows the vector field 
$(d \phi/d \tau, \, d\theta/d\tau)$ given by eq.(\ref{polar}) near 
the fixed points.  From this figure, we find that most points in 
this region go to the outside of the region or to 
the fixed point ${\bf c}_4$. 

%%%%%%%%%%%%%% 
%   Fig: vctfld
%%%%%%%%%%%%%%
%%%%%%%%%%%%%end of Fig:vctfld%%%%%%%%%%%%

The
destination of a trajectory toward the outside of this region 
is a point on the curve $\ba \cdot \bV(\ba) =0$.  This implies that 
a flow of the original RGE (\ref{orgex3}) corresponding to this flow on $S$
approaches the origin once and escapes from it.  Namely, those points belong 
to a massive phase.  On the other hand,  a flow that terminates at  
the stable fixed point ${\bf c}_4$ corresponds to 
a flow of the original RGE absorbed into the origin along 
the incoming straight flow line. The set of those points 
moving toward the fixed point ${\bf c}_4$ 
is in a massless phase.

In general, we have two cases corresponding to
massive and massless phases described above. 
However, there are exceptional points which 
lie on the solutions starting at one of the 
twice unstable fixed points ${\bf c}_1$, ${\bf c}_2$, ${\bf c}_5$
and arriving at one of the once unstable fixed points ${\bf c}_3$, 
${\bf c}_{\pm}$.  These exceptional solutions correspond 
to critical surfaces  
that give the phase boundary determined in the original RGE.   
More precisely,  the set of points on the exceptional trajectories  
is the intersection between a phase boundary 
determined by the original RGE and the sphere $S$. It should be noted 
that the phase boundary in the coupling space $\{ (g_1, g_2, g_3) \}$ 
forms a conical surface because $\bV(\bg)$ is homogeneous. 

%%%%%%%%%%%%%% 
%   Fig: pbd
%%%%%%%%%%%%%%
\begin{figure}[ht]
\setlength{\unitlength}{1mm}
 \begin{picture}(150, 50)(-15,0)
        \put(30,0){\epsfxsize=8cm \epsfbox{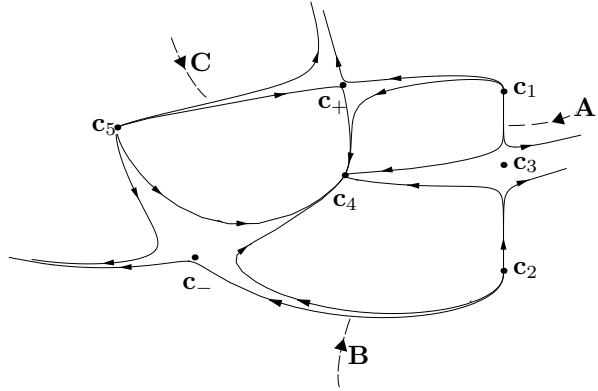}}
        \put(99,40){${\bf c}_1$}
        \put(99,30){${\bf c}_3$}
        \put(99,16){${\bf c}_2$}
        \put(55,14){${\bf c}_-$}
        \put(75,25){${\bf c}_4$}
        \put(43,35){${\bf c}_5$}
        \put(73,38){${\bf c}_+$}
                \put(107,37){${\bf A}$}
                \put(77,4){${\bf B}$}
                \put(56,43){${\bf C}$}
\end{picture}
\caption{Solutions  
 numerically computed near the exceptional ones
}
\label{pbd}
\end{figure}
%%%%%%%%%%%%%end of Fig:pbd%%%%%%%%%%%%

Fig.\ref{pbd} depicts a flow numerically solved near 
the critical surfaces.  Using the numerical results 
Figs.\ref{vctfld} and \ref{pbd}, 
let us  consider divergent behavior of the correlation length. 
Suppose that an initial value $\ba_0$ changes toward the phase boundary 
between ${\bf c}_1$ and ${\bf c}_2$ such as the dashed line 
{\bf A}.
 As $\ba_0$ tends to the phase boundary,  the flow starting at $\ba_0$ 
passes near the fixed point ${\bf c}_3$ and spends a long time there. 
The result eq.(\ref{expnt}) indicates that the critical exponent $\sigma$ 
detected in this case is 3/5. Similarly, if $\ba_0$ varies along lines 
such as {\bf B} or {\bf C}, one observes $\sigma = 8/(-4 + \sqrt{631})$ 
since the solution starting at $\ba_0$ goes through a neighborhood of 
the fixed points ${\bf c}_+$ or ${\bf c}_-$. 

Finally, we comment on the logarithmic dependence of the coupling
 constants in this example. 
Fig.\ref{vctfld} implies that a general massless  flow is controlled 
by the twice stable fixed point ${\bf c}_4$.  The universal features 
considered 
in sec.\ref{massless} are valid if the initial value $\ba_0$ is sufficiently 
far from the phase boundary.  However,  as $\ba_0$ approaches 
the phase boundary from a massless side,  the solution
is gradually affected by
the once unstable fixed points.  More profound investigation will be needed in this case.

\section{Summary and Discussion}
\label{discussion}
In this paper,  we  first showed an algebraic way of 
finding the critical exponent $\sigma$ in eq.(\ref{xi}),  which was 
so far computed by integrating RGE explicitly.  The procedure is 
summarized as follows: \\

\noindent
(i) \ Derive the new RGE defined in eq.(\ref{rgrg}) 
from the original RGE (\ref{orge2}). 

\noindent
(ii) \ Find straight flow lines in the original RGE, which correspond 
to fixed points of the new RGE. 
%The singularity of spending time as a function of the initial condition
%appears only at those fixed points.

\noindent
(iii) \ Compute the scaling matrix  at a fixed point 
on an incoming straight flow line and diagonalize it.  

\noindent
(iv) \ If the scaling matrix has the unique relevant mode, the correlation 
length indicates singular behavior by one-parameter fine tuning 
and the exponent $\sigma$ is equal to the inverse of the 
relevant eigenvalue. If the scaling matrix has multiple relevant modes, 
we can observe multicritical behavior. \\ 

Second, we derived the logarithmic dependence of running 
coupling constants in a massless phase where all the 
coupling constants are marginally irrelevant.  It was found that 
the coefficient of the leading log term is universal in the sense 
that it does not depend on an initial value of the running coupling 
constants, which is the same result as in the case of a single 
marginally irrelevant coupling constant.  However, coefficients 
of subleading terms are non-universal.  They could disturb the universal 
nature of subleading terms which come from higher-loop corrections 
 to  the original beta function.   

We obtain both results by applying the RG transformation
 (\ref{rgt}) to the original RGE (\ref{orge2}), which was motivated by 
the recent developments of RG transformations
 to non-linear differential equations\cite{BK,CGO,KHA} . 

\begin{figure}[ht]
\setlength{\unitlength}{1mm}
 \begin{picture}(150, 60)(-20,0)
        \put(30,5){\epsfxsize=8cm \epsfbox{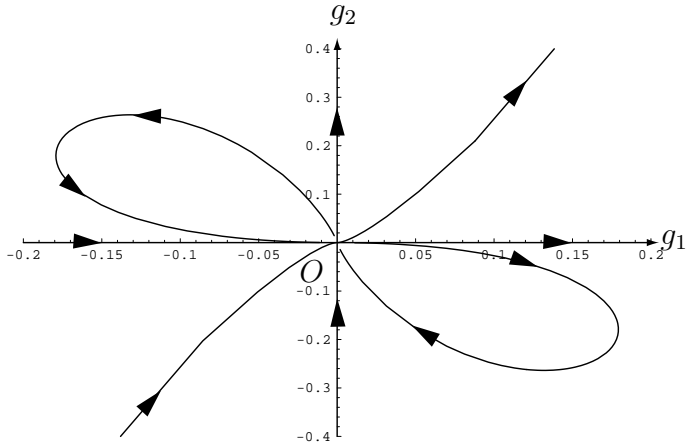}}
        \put(66,58){\large $g_2$}
        \put(62,23){\large $O$}
        \put(110,28){\large $g_1$}
\end{picture}
\caption{Flow in eq.(\ref{p=1rge}) with $q=2$
}
\label{p=1}
\end{figure}

It should be noted that our study 
is focused when we derive the first result on a flow that goes once toward 
the origin and then leaves for a non-perturbative region.   
In general, the quadratic 
differential equation (\ref{orge2}) with eq.(\ref{gigj}) 
could have a flow qualitatively different from those we considered. 
For example, we numerically find that the equation 
\beqa
        \left(
        \begin{array}{c}
                \frac{d g_1}{d t}  \\
                \frac{d g_2}{d t} 
        \end{array}
        \right) 
        = \bV(\bg) \equiv
        \left(
        \begin{array}{c}
        g_1(g_1 + g_2) \\
        g_2 (q \, g_1 + g_2)
        \end{array}
        \right)
\label{p=1rge}
\eeqa
has a flow such as in Fig.{\ref{p=1}} if $q >1$. 

%%%%%%%%%%%%%% 
%   Fig: p=1
%%%%%%%%%%%%%%
%%%%%%%%%%%%%end of Fig:p=1%%%%%%%%%%%%
Namely, the equation has solutions that first escape from the origin and 
then turn back to it.  In this case we cannot directly apply the result 
eq.(\ref{cef}).   Even in that case,  as we show in the following, 
 the beta function in eq.(\ref{rgrg}) is helpful for understanding 
a qualitative picture of these solutions.
In fact,  we first notice that there are only four straight flow lines 
on $g_1 =0$ or on $g_2 =0$.  Next let us compute 
the scaling matrix in eq.(\ref{scmt}) at the fixed points 
$(\pm a_0, 0)$ and $(0, \pm a_0)$ of the new RGE (\ref{rgrg}).   
We find that it has eigenvalues $1-q$ at $(\pm a_0, 0)$ and 
$0$ at $(0, \pm a_0)$.  If $q > 1$, solutions for
 the RGE near $(\pm a_0, 0)$ 
is absorbed into the fixed points.  The integral curve does not 
change the direction 
at the other fixed points $(0, \pm a_0)$ because the eigenvalue of the 
scaling matrix 
vanishes at those points.  Therefore, by continuity,  we conclude that 
there must be at least two branch points on $S$.  
%%%%%%%%%%%%%% 
%   Fig: p=1nrg
%%%%%%%%%%%%%%
\begin{figure}[ht]
\setlength{\unitlength}{1mm}
 \begin{picture}(150, 60)(-20,0)
        \put(30,5){\epsfxsize=8cm \epsfbox{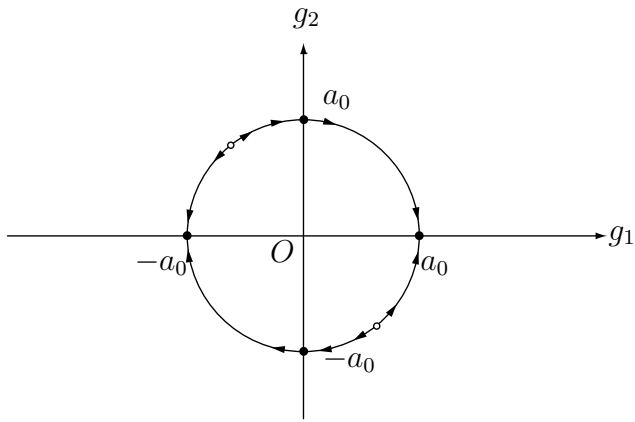}}
        \put(68,58){\large $g_2$}
        \put(85,25){\large $a_0$}
        \put(47,25){\large $-a_0$}
        \put(72,47){\large $a_0$}
        \put(72,12){\large $-a_0$}
        \put(65,26){\large $O$}
        \put(110,29){\large $g_1$}
\end{picture}
\caption{Flow in the new RGE for eq.(\ref{p=1rge}) with $q=2$. 
The white circles correspond to turning points. 
}
\label{p=1nrg}
\end{figure}
%%%%%%%%%%%%%end of Fig: p=1nrg%%%%%%%%%%%%

Moreover, the solution  
must escape from the branch points.  As we pointed out 
in the last part of \ref{formalism},  it means that 
eq.(\ref{p=1rge})  
has a flow that first leaves from the origin and turns back to 
it, as depicted in Fig. \ref{p=1}. 

The study of the universality classification
of infinite-order phase transitions 
is now in  progress. In two dimensions, 
it should contribute to the analysis of 
the $c > 1$ CFT.  In higher than two dimensions,
a nontrivial exponent in an infinite-order phase transition
might be observed experimentally in some phenomena. \\

{\bf Acknowledgments}

One of the authors (C.I) thanks J. Zinn-Justin for kind hospitality 
at Service de Physique Theorique
CEA Saclay extended to him, when part of this work was done.
He is grateful also to I. Affleck for kind hospitality at the University
of British Columbia extended to him, where this work was finished.
The authors thank J.-S. Caux to read the manuscript.
They are grateful to H. Tasaki for helpful comments.
\newpage
\appendix
\section{Relationship between the eigenvalues of $A(\ba^*)$ and 
$B(\ba^*)$}
\label{AandB.app} 
In this appendix we prove eq.(\ref{AandB}).  The claim is that 
a set of the eigenvalues of $B(\ba^*)$ in cartesian coordinates
is equal to that of $A(\ba^*)$ in polar coordinates 
plus extra zero eigenvalue. 

To show that,  we add $\ba /a \equiv \tilde{\bee}_{n}$  to the basis 
$\{\tilde{\bee}_{\al} \}_{1 \leq \al \leq n-1}$ of the tangent space 
at $\ba(\tau) \in S $.  Then the set $\{ \tilde{\bee}_i \}_{1\leq i \leq n}$
becomes an orthonormal basis of the $n$-dimensional space of coupling 
constants. 
Consider the $n \times n$ matrix 
\beq
        T_{i  j}(\ba) \equiv f_j (\ba)  (\bee_i, \tilde{\bee}_j), 
\eeq
where $\{ \bee_i \}_{1\leq i \leq n}$ 
is the orthonormal basis that defines the cartesian 
coordinates $(g_1, \cdots, g_n)$ and the bracket $(\bbox{x}, \bbox{y})$ means 
the inner product.
The function $f_j(\ba)$ is given in 
eq.(\ref{tildee}) for $1\leq j \leq n-1$.  In addition, 
\beq
        f_n (\ba) \equiv \left| \frac{\del \ba}{\del a}\right| = 1.  
\eeq
Since $(\bee_i, \tilde{\bee}_j)$ forms an orthogonal matrix, 
we immediately have the inverse of $T$ as 
\beq
        T^{-1}_{i k}(\ba) = f_i(\ba)^{-1}(\tilde{\bee}_i,  \bee_k). 
\eeq
Now we examine the form of the  $n \times n$ matrix $T^{-1} B T$. 
As the first step, let us compute $T^{-1} B$: 
\beq
        \sum_{k=1}^n 
         T^{-1}_{i k}(\ba^*) B_{k l} (\ba^*)
        = \sum_{k=1}^n f_i(\ba^*)^{-1}
        (\tilde{\bee}_i,  \bee_k) \frac{\del \be_k}{\del a_l}(\ba^*)
        = f_i(\ba^*)^{-1}
        \frac{\del }{\del a_l} (\tilde{\bee}_l,  \bbe)
        =  f_i(\ba^*)^{-1} \frac{\del  \tilde{\be}_i}{\del a_l}(\ba^*). 
\label{step1}
\eeq
Here we have used $\bbe(\ba^*)= {\bf 0}$ in the second equality. 
Next, according to eq.(\ref{tildee}), we find that $T_{i j}$ can 
be written as    
\beqa
        T_{l j} &=&  (\bee_l,  \frac{\del \ba}{\del \theta_j}) = 
        \frac{\del a_l}{\del \theta_j} \ \ ( 1 \leq j \leq n-1).   
\label{step2}
\eeqa
Using eqs.(\ref{step1}) and (\ref{step2}), we get  
\beq
        \sum_{1\leq k, l \leq n} T^{-1}_{i k}(\ba^*) 
        B_{k l}(\ba^*) T_{l j}(\ba^*) = 
        \sum_{1\leq k, l \leq n} 
        f_j(\ba^*)^{-1} \frac{\del  \tilde{\be}_i}{\del a_l}(\ba^*)
        \frac{\del a_l}{\del \theta_j}(\ba^*) = 
        f_j(\ba^*)^{-1}\frac{\del  \tilde{\be}_i}{\del \theta_j}(\ba^*)
\eeq
for $1 \leq i \leq n$ and $1 \leq j \leq n-1$.  

The result indicates that  $T^{-1} B T$ has the following form: 
\beq
          \left( T^{-1} B T \right) (\ba^*) = 
        \left(
        \begin{array}{ccc|c}
                & & & \\
                & {A(\ba^*) } & &
                 { * }\\
                & & & \\
                \hline 
                0&\cdots&0&0
        \end{array}
        \right),  
\eeq
which proves eq.(\ref{AandB}). 
Note that the last row vanishes because 
$\tilde{\be}_n(\ba) = (\bbe(\ba), \tilde{\bee}_n) = 0$ for all $\ba$.

\end{document}